\documentclass[12pt]{iopart}
\usepackage{amssymb}
\usepackage{epsfig,color}
\usepackage[latin9]{inputenc}
\usepackage{textcomp}
\usepackage{amstext}
\usepackage{amssymb}
\usepackage{graphicx}

\begin{document}
\title{Binary matter-wave compactons induced by inter-species scattering length modulations}
\author{F. Kh. Abdullaev$^{1,2}$\footnote{{Corresponding author: fatkhulla@iium.edu.my}}, M.S.A. Hadi$^{1}$, Mario Salerno$^{3}$ and B.A. Umarov$^{1}$
}
\address{
$^1$ Department of Physics, Kulliyyah of Science, International Islamic University Malaysia, 25200,  Kuantan, Malaysia.\\
$^2$ CCNH, Universidade Federal do ABC, 09210-170, Santo Andr\'e, Brazil.\\
$^3$ Dipartimento di Fisica ``E.R. Caianiello'', CNISM and INFN
- Gruppo Collegato di Salerno, Universit\'a di Salerno, Via Giovanni
Paolo II, 84084 Fisciano (SA), Italy.\\
}
\date{\today}

\begin{abstract}
Binary mixtures of quasi one-dimensional Bose-Einstein condensates
(BEC) trapped in deep optical lattices (OL) in the presence of periodic time
modulations of the inter-species scattering length, are investigated. We adopt a mean field description and use  the tight binding approximation and the  averaging method to
derive averaged model equations in the form of two coupled discrete nonlinear Schr\"odinger
equations (DNLSE) with tunneling constants that nonlinearly depend on the inter-species coupling.
We show that for strong and rapid modulations of the inter-species scattering length, the averaged system admits exact compacton solutions, e.g. solutions that have no tails and are fully localized on a compact  which are achieved when the densities at the compact edges are in correspondence with zeros of the Bessel function (zero tunneling condition). Deviations from  exact conditions  give rise to the formation of quasi-compactons, e.g. non exact excitations which look as compactons for  any practical purpose, for which the zero tunneling condition is achieved dynamically thanks to an effective nonlinear dispersive coupling induced by the scattering length modulation.  Stability properties of compactons and  quasi-compactons are investigated by linear analysis and by numerical integrations of the averaged system, respectively, and results compared with those from the original (unaveraged) system. In particular, the occurrence od delocalizing transitions with existence of thresholds in the mean inter-species scattering length  is explicitly demonstrated. Under
proper management conditions, stationary compactons and quasi-compactons are quite stable and
robust excitations that can survive on very long time scale. A parameter design and a possible experimental setting  for observation of these excitations  are briefly discussed.
\end{abstract}

\maketitle

\section{Introduction}

Localized monlinear waves with a compact support, so called compactons, are recently attracting interest mainly from mathematical and theoretical points of view. Their existence was
predicted for equations of the KdV type with nonlinear
dispersion~\cite{Rosenau1} and later demonstrated for several continuous~\cite{Rosenau2} and discrete nonlinear systems in one and two spatial
dimensions~\cite{Pikovsky,Johansson}. In spite of the many
investigations, however, they  are presently still unobserved in
physical experiments. From this point of view appears particularly interesting
the theoretical work done on compactons of the discrete nonlinear Schr\"odinger
equation (DNLSE), a model that is linked to physical
systems such as nonlinear optical waveguides arrays and Bose-Einstein
condensates (BEC) trapped in deep optical lattices (OL). In these contexts,
it has been demonstrated that when subjected to periodic time dependent rapid modulations
of the nonlinearity, also called strong
nonlinearity management (SNM) limit, the DNLSE can support stable compactons of different
types~\cite{AKS,AHSU,DSKA}. In particular, the existence of one-dimensional compactons
in BEC arrays~\cite{AKS} and in binary BEC mixtures with time modulated
{\it intra-spices} scattering length~\cite{AHSU}, as well as, the existence of compactons and vortex-compactons (e.g. vortices on a compact support)  in two-dimensional DNLSE~\cite{DSKA}, were recently demonstrated.

From a physical point of view the existence of compactons is a direct
consequence of the tunneling suppression across the edges of the excitation.
This phenomenon originates from the  nonlinear
dispersion term  which introduces a dependence of
the  tunneling  constant on the density  imbalance  between
adjacent sites, so that for imbalances matching  proper values the
tunneling becomes locally suppressed~\cite{AK,GMH}.
Experimentally, this phenomenon  has been recently observed in BEC in deep OL with periodic time modulatations of the interactions~\cite{Meinert}. We also remark that the SNM  of the scattering lengths is presently  used  to investigate several interesting physical phenomena related to new quantum phases~\cite{Greschner,Akos}, artificial
gauge fields and spin-orbit couplings~\cite{Wang,GSP}, in BEC systems.

The aim of the present paper is to study binary mixtures of quasi
one-dimensional Bose-Einstein condensates in deep optical lattices
in the presence of periodic modulations of the \textit{inter-species}
scattering length. In this respect, we derive an effective fast time-averaged
vector DNLS equation and show that compactons obtained in this
case depend on a nonlinear rescaling of the tunneling constants that
involves directly the coupling of the two components. In the SNM limit
this leads to stable binary compacton solutions with different properties from the ones obtained
under an \textit{intra-species} management of the scattering length.
More specifically, we show that in addition to the stationary compactons
achieved when the edge amplitudes are in correspondence with zeros of the Bessel function (zero tunneling condition), there exist also stationary quasi-compactons for
which the zero tunneling condition is achieved dynamically, via the effective coupling induced by the interspecies
scattering length modulation. These solutions exist and are stable on a long time scale especially when one of the two components (say the first)  matches the zero tunneling condition in the uncoupled limit, while the other slightly deviates from it. Quite interestingly, we find that
for fixed deviations of the amplitude of the second component from a zero
of the Bessel function, there exists a threshold for the mean inter-species
scattering length above which the zero tunneling condition can be dynamically
established for both components and stationary quasi-compactons can exist.
For inter-species scattering lengths below the existence threshold, the component matching the zero tunneling condition (in the uncoupled limit) remains alive while the other  quickly decays to zero, this leading to the destruction of the binary compacton. As the deviation, for fixed system parameters, from the zero tunneling condition of the inexact component is reduced, the existence threshold moves toward smaller values and completely disappear when the deviation become smaller that a critical value. In the last case it becomes possible to have long living stationary quasi-compactons even if the mean inter-species scattering length is detuned to zero.
The solutions predicted by our analysis are checked by means of direct
numerical simulations of the full vector DNLSE with inter-species scattering
length management. The comparison between analytical and numerical
results shows a very good agreement in all the investigated cases.

The paper is organized as follows. In Sec.~2 we introduce the
model equations for binary a BEC mixtures and use the averaging method
to derive the effective equations valid in the limit of strong modulations
of the interspecies scattering length. In Sec.~3 we use the averaging
equations to investigate the existence and the stability of exact stationary
bright-bright compactons localized on one, two and three
lattice sites and compare results with the ones obtained from direct
numerical integrations of the original (unaveraged) system. In Sec.~4
we consider stationary quasi-compactons for which the condition of
the zero tunneling at the compacton edges is dynamically achieved
through the inter-species coupling of the BEC mixture. In the last
section we discuss physical parameters for possible experimental observations
and briefly summarize the main results of the paper.

\section{Model equations and averaging}
The model equations for a  BEC mixture in a one-dimensional (1D) geometry can be derived from a more general three-dimensional
formalism by considering a trapping potential with the transversal frequency $\omega_\perp$ much larger than the longitudinal frequency $\omega_{||}$, e.g.
$\omega_\perp\gg\omega_{||}$. In the present case, the trap potential in the $x-$direction is an OL generated by two counter-propagating laser fields giving rise to a periodic standing wave potential of the form $V_{ol}(x)\equiv V \cos(2k_L x)$ with $k_L$ denoting the optical lattice wave-number. In the mean field approximation, the system is described by a
1D Gross-Pitaevskii (GP) coupled equation for the two-component wave function, $\Psi_j\equiv\Psi_j (x,t), \ j=1,2$,
\begin{eqnarray}
i\hbar\frac{\partial \Psi_{j}}{\partial t} & =&\left[
-\frac{\hbar^2}{2m_j}\frac{\partial^2}{\partial x^2 }
+V_{ol}(x) \right]\Psi_j + 2\hbar\omega_\perp(a_{jj}|\Psi_j|^2 + a_{j,3-j}|\Psi_{3-j}|^2)\Psi_j
\;\;\;
\label{eqGP}
\end{eqnarray}
where  $a_{jj}\;\;(j=1,2)$ and $a_{12}$ are the two-body scattering lengths between
intra- and inter-species of atoms. Eq.~(\ref{eqGP}) can be put in dimensionless form by performing the following change of variables
\begin{equation}
x\to\frac{x}{k_L},\;\; t\to \frac{\hbar}{E_R}\, t, \;\; V\to \frac{V}{E_R}\equiv V_0, \;\;
\end{equation}
where space is measured in units of $1/k_L$, the energy in units of the lattice recoil energy $E_R \equiv \hbar \omega_R= \hbar^2 k_L^2/2 \mu$ with  $\mu=m_1 m_2/(m_1+m_2)$ the reduced mass,  and time in units of $\hbar/E_R$, this leading to
\begin{eqnarray}
i \frac{\partial \psi_{j}}{\partial t} & =& \left[- \epsilon_j \frac{\partial^2}{\partial x^2} + V_0 \cos (2 x) \right] \psi_{j} +  (g_j |\psi_{j}|^2+ {g} |\psi_{3-j}|^2) \psi_j, \;\;\;j=1,2.
\label{normGPE}
\end{eqnarray}
In the above equation the wavefunctions $\Psi_j$ were rescaled according to
\begin{equation}
\Psi_j\equiv\sqrt{\frac{\omega_R}{2\omega_{\perp}a_0}}  \psi_j(x,t)
\end{equation}
and we denoted $\epsilon_j= \mu/m_j, \; j=1,2,$ $g_j\equiv\frac{a_{jj}}{a_0},\;\; g\equiv\frac{a_{j,3-j}}{a_0}$, with $a_0$ the background scattering length.

For deep optical lattices, e.g. when the OL amplitude is relatively large compared to $E_R$ (say $V > 10 E_R$ or  $V_0 \gg 1$), it is possible to make the tight-binding approximation~\cite{TS,ABKS} by expanding the two component wavefunctions $\psi_j$ in terms of Wannier functions of the underlying linear periodic problem~\cite{AKKS}
\begin{equation}
\psi_1(x,t)= \sum_n u_n(t) W_1(x-x_n),\;\;\; \psi_2(x,t)= \sum_n v_n(t) W_2(x-x_n),
\label{Wannexp}
\end{equation}
with time dependent expansion coefficients $u_n, v_n$ to be fixed in such a manner that Eq.~(\ref{normGPE}) is satisfied. By substituting the above expansions in Eq.~(\ref{normGPE}) and using the orthonormality property of the Wannier functions $\int dx W_{j}(x-n)^* W_{j'}(x-n')=\delta_{j,j'} \delta_{n,n'}$, one gets the following coupled discrete nonlinear Schr\"odinger  equations (DNLSE)~\cite{Shrestha}
\begin{eqnarray}
 && i\dot{u}_{n}=-\kappa_{1}(u_{n+1}+u_{n-1})-(\gamma_{1}|u_{n}|^{2}+\gamma_{12}|v_{n}|^{2})u_{n},\nonumber \\
\label{vdnlse}
\\
 && i\dot{v}_{n}=-\kappa_{2}(v_{n+1}+v_{n-1})-(\gamma_{12}|u_{n}|^{2}+\gamma_{2}|v_{n}|^{2})v_{n},\nonumber
 \end{eqnarray}
where the overdot stands for time derivative. The coefficients $\kappa_{i},\;i=1,2$, related to the tunneling rates of atoms between neighboring OL potential wells, and coefficients  $\gamma_{j},\;j=1,2$, related  to the effective inter-species and
intra-species nonlinearities, respectively, are expressed in terms of overlap integrals of Wannier functions as:
\begin{eqnarray}
& & \kappa_j = \epsilon_j \int W_{j}^*(x-n)\frac{\partial^2}{\partial x^2}W_{j}(x-(n+1)) dx, \nonumber \\
& & \gamma_{j}=
g_j \int dx |W_{j}(x-n)|^{4},\;\;\;\;\;j=1,2, \\
&  &
\gamma_{12}=
g \int dx|W_{1}(x-n)|^{2}|W_{2}(x-n)|^{2}. \nonumber
\end{eqnarray}
Since the same OL acts on both components, the underlying linear periodic system is the same and so are the Wannier functions and corresponding tunneling coefficients, e.g.  $\kappa_1=\kappa_2\equiv \kappa$.

Notice that Eq.~(\ref{vdnlse})  has the Hamiltonian form $\dot{\chi}_{n}=\delta H/\delta\chi_{n}^{*}$
with $\chi_{n}=u_{n},v_{n}$ and Hamiltonian $H$ given by
\begin{equation}\hskip -1.8cm
H=-\sum_{n}\Big[(\kappa_{1}u_{n+1}u_{n}^{\ast}+\kappa_{2}v_{n+1}v_{n}^{\ast}+c.c.)\,+
  \frac{1}{2}\left(\gamma_{1}|u_{n}|^{4}+\gamma_{2}|v_{n}|^{4}\right)+\gamma_{12}|u_{n}|^{2}|v_{n}|^{2}\Big].\label{HOri}
\end{equation}
Here the star $"*"$  stands for complex conjugation and $c.c.$ denotes
the complex conjugate of the expression in the parenthesis. In addition to the Hamiltonian (energy),  the numbers of atoms $N_{i}=\sum_{n}|\chi_{n}|^{2},\;\;\chi_{n}=u_{n},v_{n}$ in each component
are also conserved quantities.
It is interesting to remark here  that Eq.~(\ref{vdnlse}) also arises in nonlinear optics in connection with the propagation of the electric field in an array of optical waveguides
with variable Kerr nonlinearity. In this context the role of time
is played by the longitudinal propagation distance along the optical
fiber and the nonlinear coefficients $\gamma_{ij}$ correspond to
self- and cross-phase modulations of the electric field components,
respectively~\cite{Kobyakov,Assanto}.

In the following we consider BEC mixtures with fixed intra-species nonlinearities and
subjected to periodic time modulations of the inter-species  nonlinear parameter $\gamma_{12}$
of the form
\begin{equation}
\gamma_{12}(t)\equiv\gamma_{12}^{(0)}+\gamma_{12}^{(1)}(t)=\gamma_{12}^{(0)}+\frac{\gamma_{12}^{(1)}}{\epsilon}\cos(\Omega\frac{t}{\epsilon}),\;\;\;\ \label{modulation}
\end{equation}
with $\gamma_{12}^{(0)},\gamma_{12}^{(1)}$ real constants and $\varepsilon$
a small parameter controlling the strongness of the modulation (SNM  corresponds to $\epsilon\ll1$ with $\Omega,\;\gamma_{12}^{(1)}\sim O(1)$).
We remark that the choice in Eq.~(\ref{modulation}) of an harmonic
modulating function of period $T=2\pi/\Omega$ in the fast time variable $\tau=t/\epsilon$ is just for analytical convenience, what is important is the  periodicity property of the modulation function rather than its shape, results being of general validity and can be easily  extended to other types of periodic functions $\gamma_{12}^{(1)}(t)$. Periodic variations of the scattering length in time
can achieved by using the Feshbach resonance technique~\cite{Inouye, REV-MOD-PHYS}.
This management setting is quite convenient in experiments since it
involves only one parameter e.g. the inter-species scattering length
(for the case of intra-species management see~\cite{AHSU}). Moreover,
the averaged equations obtained in the present case involve the coupling of the BEC components trough the nonlinear dispersive terms that leads
to novel types of binary compactons (see below).

To find effective nonlinear evolution equations, we use averaging
method to eliminate the fast time, $\tau=t/\epsilon$, dependence.
In this respect it is convenient to perform the following transformation
\begin{equation}
u_{n}=U_{n}e^{i\Gamma|V_{n}|^{2}},\;v_{n}=V_{n}e^{i\Gamma|U_{n}|^{2}},\label{eq2}
\end{equation}
where $\Gamma$ denotes the antiderivatives of $\gamma_{12}^{(1)}(t)$,
e.g. $\Gamma_{12}(\tau)=\frac{\gamma_{12}^{(1)}}{\varepsilon}\int_{0}^{\tau}\cos(\Omega\tau')d\tau'$.
By substituting Eq.~(\ref{eq2}) into Eqs.~(\ref{vdnlse}) we obtain:
\begin{eqnarray}
i\dot{U}_{n}&= & \,i\kappa_{2}\Gamma(\tau)U_{n}[V_{n}^{\ast}X_{1}-V_{n}X_{1}^{*}]-
 \kappa_{1}X_{2}-(\gamma_{1}|U_{n}|^{2}+\gamma_{12}^{0}|V_{n}|^{2})U_{n},\\
i\dot{V}_{n}&= & \,i\kappa_{1}\Gamma(\tau)V_{n}[U_{n}^{\ast}X_{2}-U_{n}X_{2}^{*}]-
 \kappa_{2}X_{1}-(\gamma_{2}|V_{n}|^{2}+\gamma_{12}^{(0)}|U_{n}|^{2})V_{n},
\end{eqnarray}\label{eq.tr}
where $X_{1}=U_{n+1}e^{i\Gamma\theta_{1}^{+}}+U_{n-1}e^{i\Gamma\theta_{1}^{-}}$,
$X_{2}=V_{n+1}e^{i\Gamma\theta_{2}^{+}}+V_{n-1}e^{i\Gamma\theta_{2}^{-}}$
and with $\theta_{i}^{\pm}$ denoting the quantities
\begin{equation}
\theta_{1}^{\pm}=|U_{n\pm1}|^{2}-|U_{n}|^{2},\;\;\;\theta_{2}^{\pm}=|V_{n\pm1}|^{2}-|V_{n}|^{2}.
\end{equation}

The average over the rapid modulation can be easily done with the help of the following relations
\begin{equation}
<e^{\pm i\Gamma(\tau)\theta^{\pm}}>  =J_{0}(\alpha\theta^{\pm}), \;\;\;
<\Gamma(\tau)e^{\pm i\Gamma(\tau)\theta^{\pm}}>=\pm i\,\alpha J_{1}(\alpha\theta^{\pm}),
\label{averages}
\end{equation}
where the angular bracket $<F>$ denotes the average with respect
to the fast time, e.g. $<F>\equiv(1/T)\int_{0}^{T}Fd\tau$, while
$J_{0}$, $J_{1}$ are Bessel functions~\cite{abramowitz} of first
kind of zero-th and first order, respectively, with the parameter $\alpha$
given by
\begin{equation}
\alpha=\gamma_{12}^{(1)}/\Omega.\label{alfa}
\end{equation}
The system of averaged equations is then obtained as:
\begin{eqnarray}
i\dot{U}_{n}=-  \alpha\kappa_{2}U_{n}\Big[J_{1}(\alpha\theta_{1}^{+})\,\left(V_{n}^{\ast}V_{n+1}+V_{n}V_{n+1}^{\ast}\right)
  +J_{1}(\alpha\theta_{1}^{-})\,\left(V_{n}^{\ast}V_{n-1}+V_{n}V_{n-1}^{\ast}\right)\Big]\nonumber\\
-  \kappa_{1}\left[U_{n+1}J_{0}(\alpha\theta_{2}^{+})+U_{n-1}J_{0}(\alpha\theta_{2}^{-})\right]
-  \left[\gamma_{1}|U_{n}|^{2}+\gamma_{12}^{(0)}|V_{n}|^{2}\right]U_{n},
\label{eq.av1}
\end{eqnarray}
\begin{eqnarray}
i\dot{V}_{n}=-  \alpha\kappa_{1}V_{n}\Big[J_{1}(\alpha\theta_{2}^{+})\,\left(U_{n}^{\ast}U_{n+1}+U_{n}U_{n+1}^{\ast}\right)
  +J_{1}(\alpha\theta_{2}^{-})\,\left(U_{n}^{\ast}U_{n-1}+U_{n}U_{n-1}^{\ast}\right)\Big]\nonumber\\
-  \kappa_{2}\left[V_{n+1}J_{0}(\alpha\theta_{1}^{+})+V_{n-1}J_{0}(\alpha\theta_{1}^{-})\right]
-  \left[\gamma_{2}|V_{n}|^{2}+\gamma_{12}^{(0)}|U_{n}|^{2}\right]V_{n}.
\label{eq.av2}
\end{eqnarray}
Note that Eqs. (\ref{eq.av1}, \ref{eq.av2}) have Hamiltonian form
with averaged Hamiltonian $H_{av}$ given by:
\begin{eqnarray}
H_{av}=- \sum_{n}\Big[\kappa_{1}J_{0}(\alpha\theta_{2}^{+})[U_{n+1}U_{n}^{\ast}+ c.c] + \kappa_{2}J_{0}(\alpha\theta_{1}^{+})\,[V_{n+1}V_{n}^{\ast}+c.c.] \nonumber\\ \qquad\qquad\qquad\qquad\qquad
 + \frac{1}{2}\left(\gamma_{1}|U_{n}|^{4}+\gamma_{2}|V_{n}|^{4}\right)+\gamma_{12}^{(0)}|U_{n}|^{2}|V_{n}|^{2}\Big].\label{Hav}
\end{eqnarray}
By comparing Eq.~(\ref{Hav}) with the corresponding
unperturbed Hamiltonian Eq.~(\ref{HOri}), one can see that the effect
of the inter-species scattering length modulation reflects  in the following nonlinear
rescaling of the tunneling constants:
\begin{equation}
\kappa_{i}\rightarrow\kappa_{i}J_{0}(\alpha\theta_{3-i}^{+}),\;\;\;\;\;\;\;i=1,2.\label{rescaling}
\end{equation}
Notice that, differently from the intra-species management considered
in~\cite{AHSU}, the tunneling constant of one species depends
on the difference of the atomic density between neighboring sites~\cite{GMH,AKS}
of the other species. This introduces nonlinear dispersive coupling
terms in the averaged equations leading  to novel types of two-component
compactons. It is also worth to note here that the appearance of the
Bessel functions in the above equations is a consequence of the choice of a sinusoidal
modulation function. Compacton existence  and
lattice tunneling suppression, however, are independent on this. One can easily show that in the general case of a periodic (not sinusoidal) modulation of the scattering lengths, the above averaged  equations
will retain the same form except for the replacement of Bessel functions of the argument $\alpha\theta_{i}^{\pm}$ with more involved functions of the same arguments.

\section{Exact binary compactons, stability analysis and numerical results}
In this section we derive exact compacton solutions and study their  stability properties both by standard linear analysis and by direct numerical integrations of the original nonlinear equations.

Exact compacton solutions of the averaged system can be searched in the form of stationary states
\begin{equation}
U_{n}=A_{n}e^{-i\mu_{u}t},\ V_{n}=B_{n}e^{-i\mu_{v}t},
\end{equation}
with $\mu_{u},\mu_{v}$ chemical potentials of the two atomic species.

By substituting these expressions into Eq.~(\ref{eq.av1},\ref{eq.av2})
one gets the following stationary equations:
\begin{eqnarray}
\mu_{u}A_{n}+(\gamma_{1}A_{n}^{3}+\gamma_{12}^{(0)}B_{n}^{2}A_{n})+\kappa_{1}[A_{n+1}J_{0}(\alpha\theta_{2}^{+})+
A_{n-1}J_{0}(\alpha\theta_{2}^{-})] \nonumber  \\ \qquad\qquad\qquad\quad\quad + 2\alpha\kappa_{2}A_{n}B_{n}[B_{n+1}J_{1}(\alpha\theta_{1}^{+})+B_{n-1}J_{1}(\alpha\theta_{1}^{-})]=0,\label{mubg1} \\
\mu_{v}B_{n}+(\gamma_{2}B_{n}^{3}+\gamma_{12}^{(0)}A_{n}^{2}B_{n})+\kappa_{2}[B_{n+1}J_{0}(\alpha\theta_{1}^{+})+
B_{n-1}J_{0}(\alpha\theta_{1}^{-})] \nonumber \\ \qquad\qquad\qquad\quad\quad + 2\alpha\kappa_{1}A_{n}B_{n}[A_{n+1}J_{1}(\alpha\theta_{2}^{+})+A_{n-1}J_{1}(\alpha\theta_{2}^{-})]=0, \label{mubg2}
\end{eqnarray}
to be solved for the chemical potentials and amplitudes $A_{n},B_{n}$
of the compacton modes.

To search for exact  compactons one must look for
the last sites of vanishing amplitude, say $n_{0}\pm1$, where the vanishing of the
tunneling rate is realized. The compact nature of the solution ($A_{i},B_{i}=0$
outside a finite small range of sites), allows to reduce the above
infinite system into a finite number of relations between the above
variables, which can be solved exactly. This is shown explicitly below for
different compacton types. For simplicity in the following we concentrate on
bright-bright (B-B) solutions of the BEC mixtures. Similar results can be derived also for bright-dark and dark-dark compactons (omitted  for brevity), in analogy to what done in Ref.~\cite{AHSU} for the intra-species SNM case.

\begin{figure}
\centerline{
\hskip -.5cm
\includegraphics[scale=0.6]{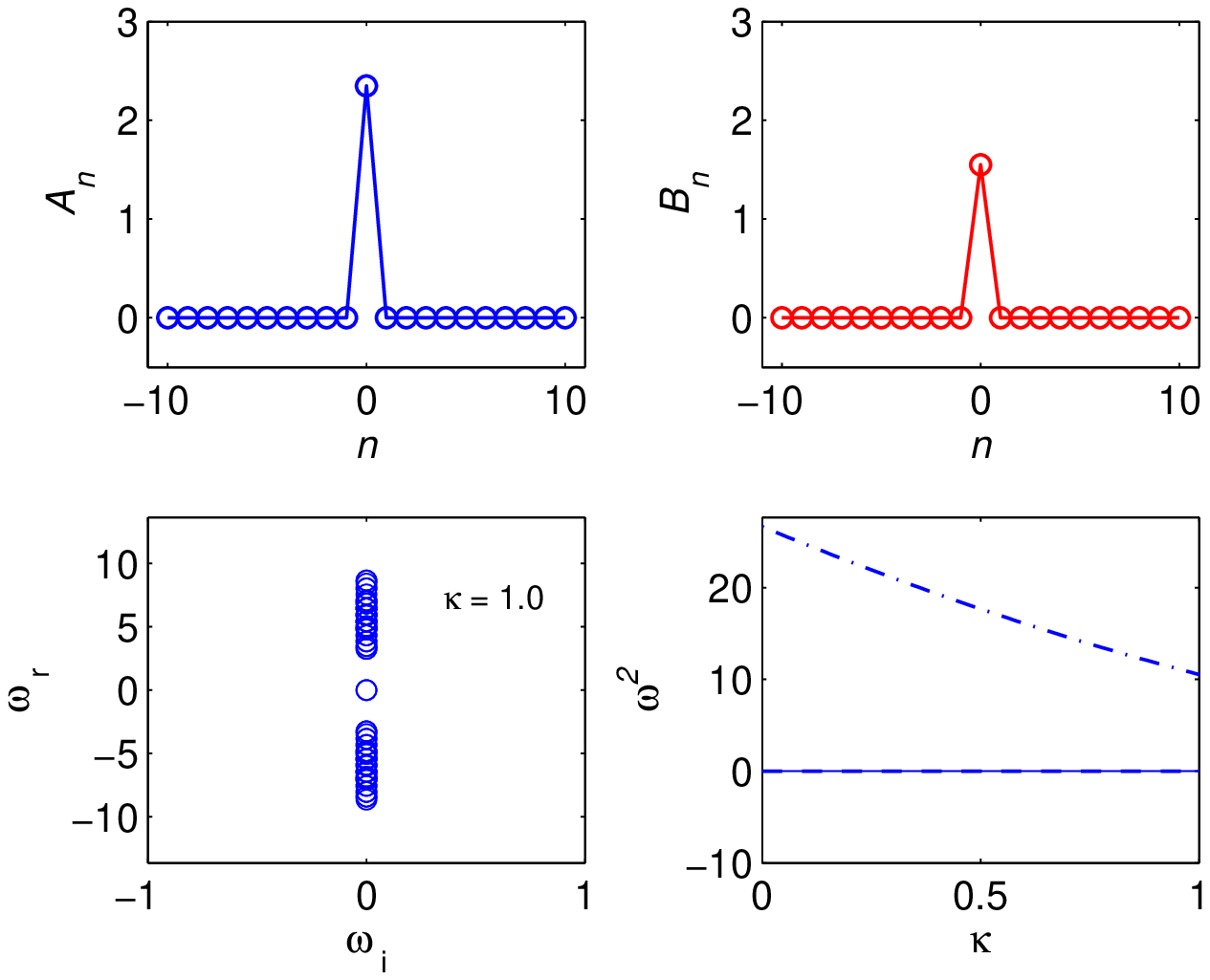}
\hskip -.9cm
\includegraphics[scale=0.6]{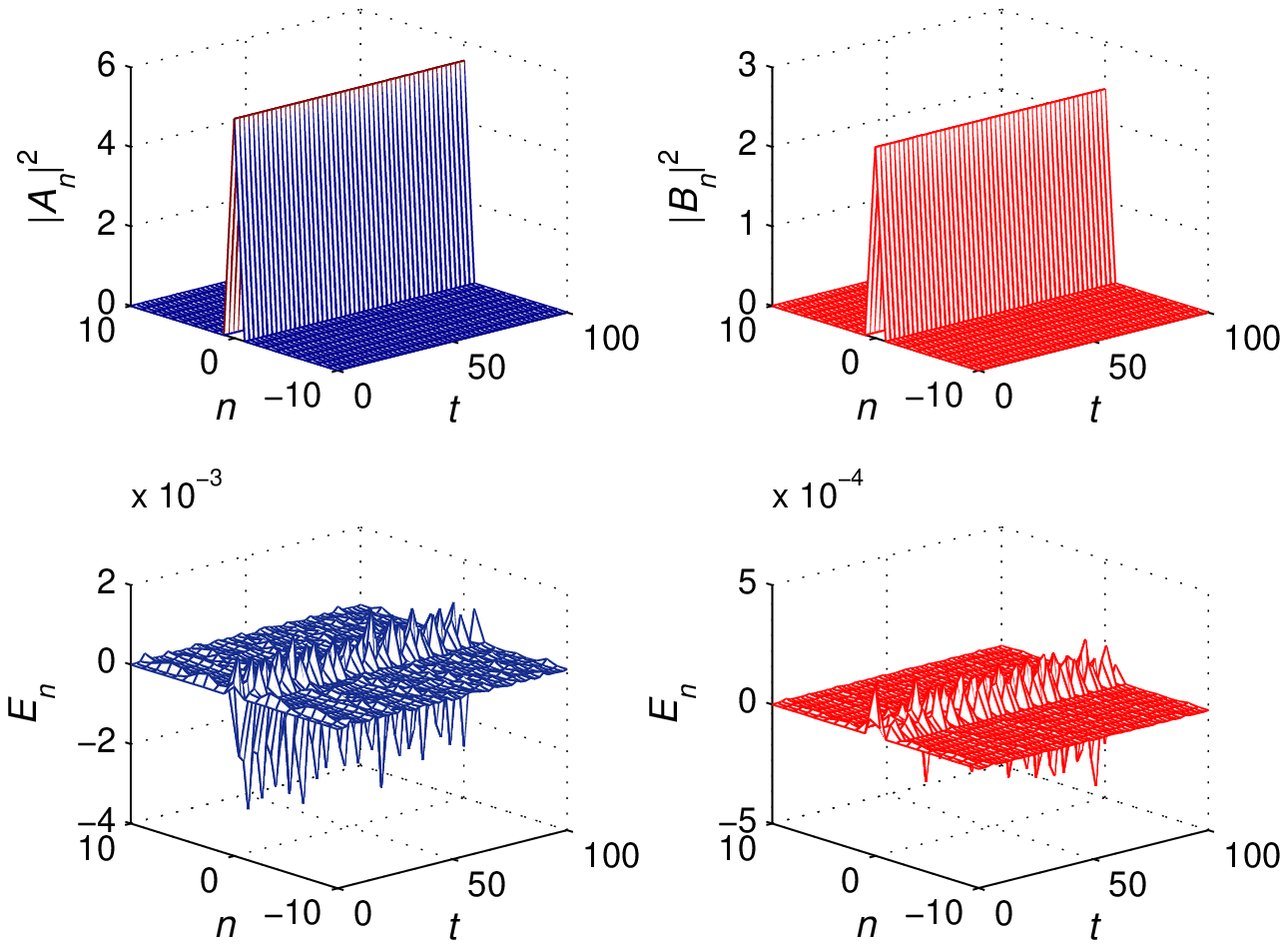}
}
\caption{
(Color online)
 \underline{Top panels}. Exact single site B-B compacton of the averaged
Eqs. (\ref{eq.av1}-\ref{eq.av2}) with first (first panel from the left) and second (second panel from the left)
component amplitude fixed as $\left|A_{0}\right|^{2}=5.0074$ (second zero of $J_0$), $\left|B_{0}\right|^{2}=2.4048$ (first zero of $J_0$), respectively. Third and fourth top panels (from the left) show the space-time evolution of $|A_{n}|^{2}$ and $|B_{n}|^{2}$, respectively, as obtained from direct numerical integration of Eq.~(\ref{vdnlse}). \underline{Bottom panels}. In the first panel (from the left) are shown the real and imaginary parts of the eigenfrequency  $\omega$ of the linear system obtained by linearizing Eqs. (\ref{eq.av1}-\ref{eq.av2})  around the B-B compacton depicted in the top left panel, while in the second panel from the left are shown the lowest (solid line), third lowest (dash line),
and fifth lowest (dash-dot line) $\omega^{2}$ values as a function
of the parameter $\kappa$. The remaining bottom panels show the time dependence of the deviations $E_n(t)\equiv |u_n(t)-u_n(0)|$, $E_n(t)\equiv |v_n(t)-v_n(0)|$,  of the first (second panel from the right) and second (first  panel from the right) component amplitudes depicted in the corresponding top panels, from their initial values at $t=0$.
In all the panels parameter values are fixed as: $\gamma_{1}=\gamma_{2}=1$, $\gamma_{12}^{(0)}=0.5$, $\kappa_{1}=\kappa_{2}\equiv\kappa=1$ and $\alpha=1$ .
The numerical integration of Eq.~(\ref{vdnlse}) is made by taking as initial condition  the exact single site B-B compacton shown in the top left panel, the modulation function
$\gamma_{12}(t)=\gamma_{12}^{(0)}+\frac{\alpha}{\epsilon}\cos\left(t/\epsilon\right)$, with
$\epsilon=0.01$ and all other parameters fixed as above.}

\label{brightbrightsingle}
\end{figure}

\subsection{One-site compactons} A single site B-B compacton is achieved by assuming the following pattern for the amplitudes
$A_{n_{0}}=a,\,B_{n_{0}}=b,\;A_{n_{0}\pm j}=0,\,B_{n_{0}\pm j}=0$
for all $j\ge1$, with a, b real constants to be determined.
Substituting this ansatz in Eqs.(\ref{mubg1},\ref{mubg2})
we obtain the corresponding conditions for the compacton existence as
\begin{eqnarray}
J_{0}(\alpha a^{2})=0,\ a^{2}=\xi_{1}/\alpha,\nonumber \\
J_{0}(\alpha b^{2})=0,\ b^{2}=\xi_{2}/\alpha\label{aconst}
\end{eqnarray}
where $\xi_{1},\xi_{2}$ are zeros (not necessary equal) of the Bessel
function $J_{0}$. This condition together with
\begin{equation}
\mu_{u}=-\gamma_{1}a^{2}-\gamma_{12}b^{2},\ \mu_{v}=-\gamma_{2}b^{2}-\gamma_{12}a^{2}
\end{equation}
gives us the single site B-B compacton pair. Typical examples of B-B compactons are shown in top panels of Fig.~\ref{brightbrightsingle}
together with their stability properties investigated by means of standard  linear analysis performed on  the averaged equation of motion. Denoting by $\omega$ the eigenfrequencies of the linearized averaged equations, associated with growing perturbation of the form $e^{-i\omega t}$, we have that linear stability is granted if all $\omega$ have zero imaginary parts. From the second row panels of  Fig.~\ref{brightbrightsingle} we see that one site B-B compactons are linearly very stable for a wide range of the parameter $\kappa$ variation. Stability has been check also by direct numerical integrations of Eq.~(\ref{vdnlse})  shown in the two bottom panels of the figure.
\begin{figure}
\hskip -.4cm
\centerline{\includegraphics[scale=0.6]{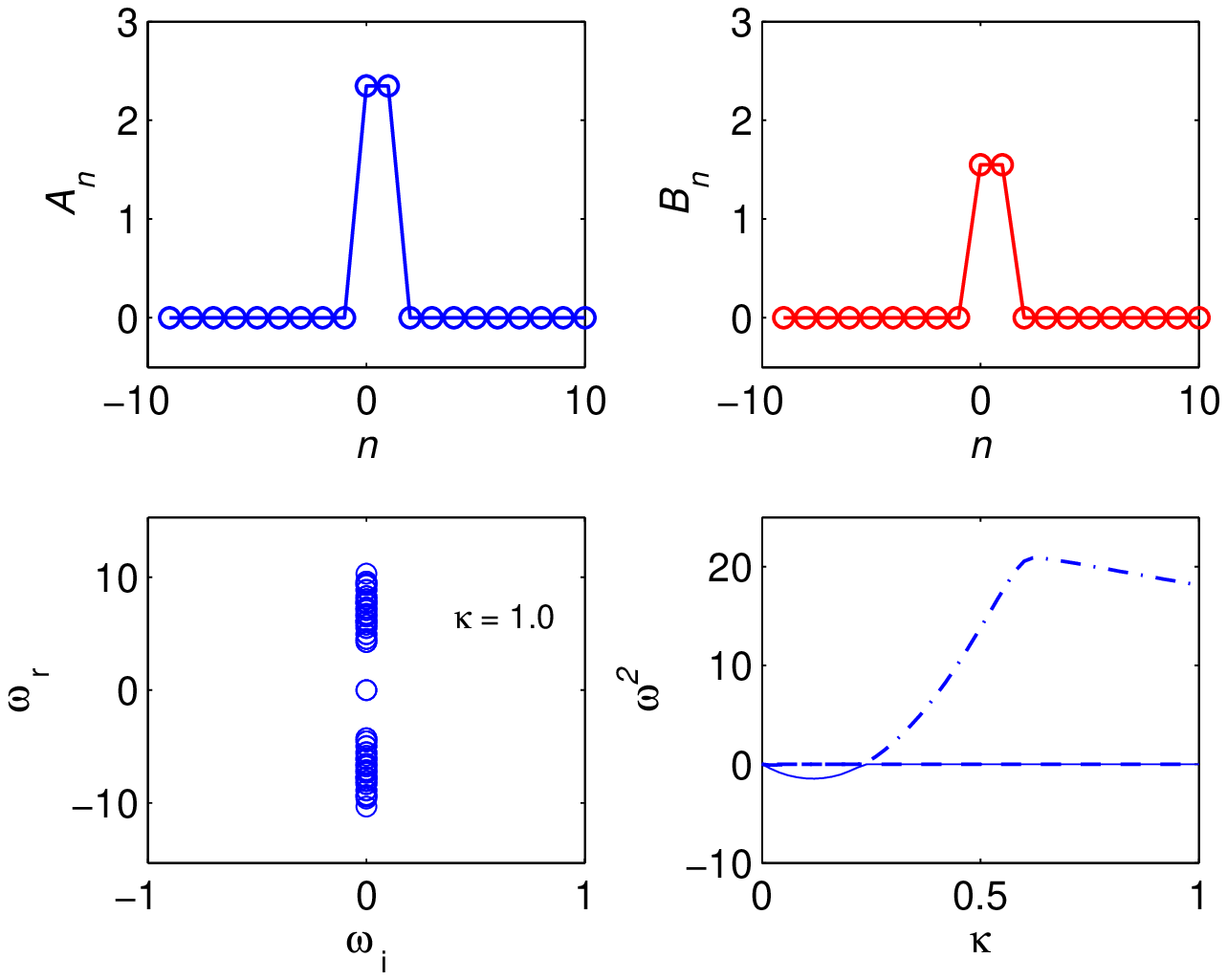}
\hskip -.8cm
\includegraphics[scale=0.6]{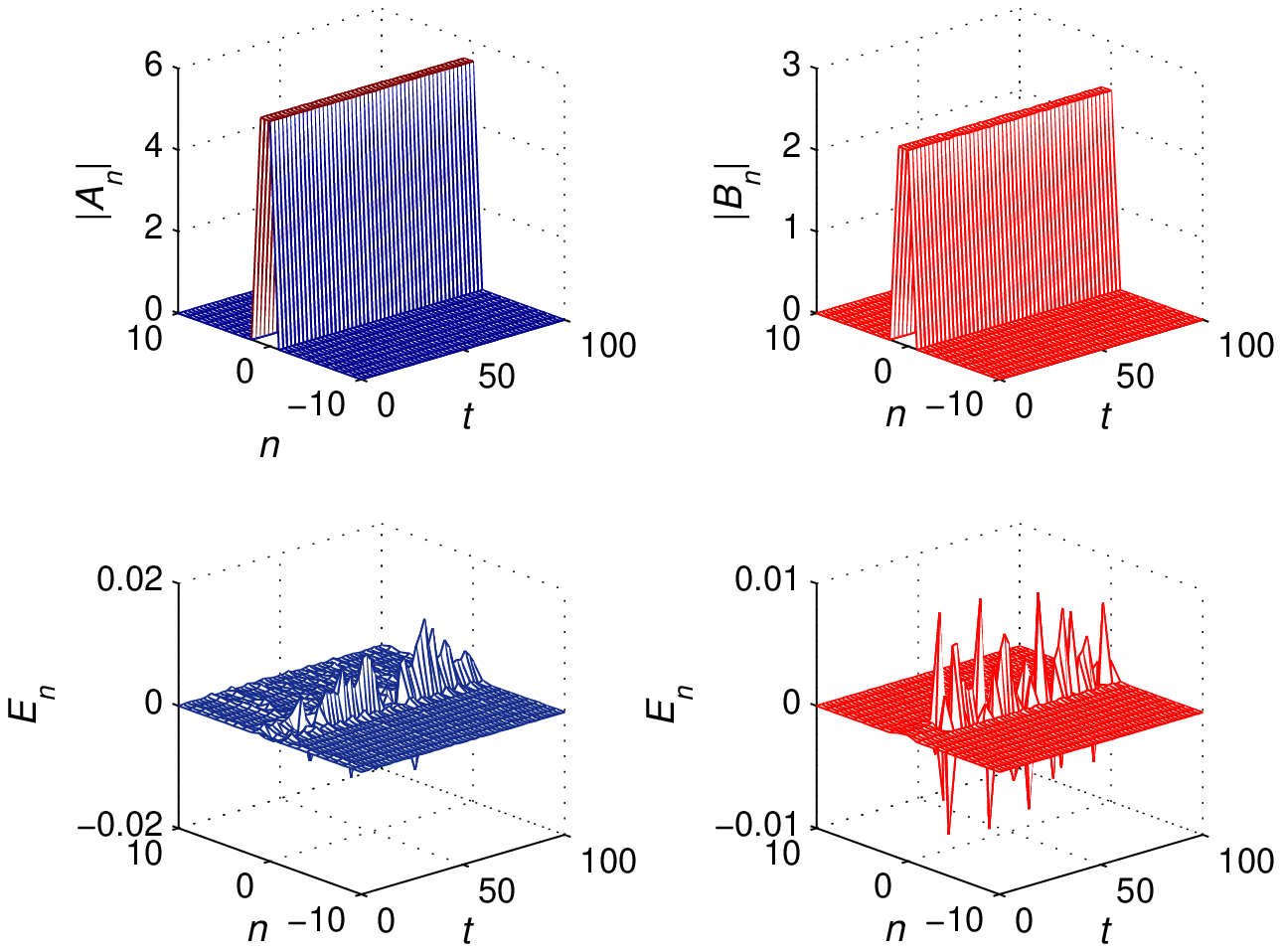}}
\protect\protect\protect\protect\caption{(Color online) Same as in Fig.~\ref{brightbrightsingle} but for a two-site in-phase B-B compacton corresponding to the amplitude pattern given in Eq.~(\ref{IS-IS}) with $a, b$ fixed as the square root of the second and first zero of the Bessel function $J_0$, respectively.}
\label{brightbrightwosite}
\end{figure}

\subsection{Two-site compactons}
Two-site B-B compactons also exist and, similarly to discrete breathers of  binary non linear lattices~\cite{CBKAS}, can be of three types: \textit{in-phase}, \textit{out-of-phase} and  \textit{mixed-symmetry} type, e.g. with both components symmetric (in-phase),  or with both  anti-symmetric (out-of phase), or with one component symmetric and the other  anti-symmetric (mixed-symmetry) ~\cite{CBKAS},  with respect to the middle point between the two sites on which they are localized, respectively. Two-site B-B compactons directly  follow from Eqs.~(\ref{mubg1}, ~\ref{mubg2}) by assuming one of the following patterns, depending on their symmetry:
\begin{eqnarray}
\label{IS-IS}
A_{n_{0}}=a,\,,\,A_{n_{0}+1}=a,\quad B_{n_{0}}= b,\,,\,B_{n+1}=b,
\\
\label{IA-IA}
A_{n_{0}}=a,\,\,A_{n_{0}+1}=-a,\quad B_{n_{0}}= b,\,\,B_{n+1}=-b,
\\
\label{IS-IA}
A_{n_{0}}=a,\,A_{n_{0}+1}=a,\quad B_{n_{0}}= b,\,\,B_{n+1}=-b,
\\
\label{IA-IS}
A_{n_{0}}=a,\,A_{n_{0}+1}=-a,\quad B_{n_{0}}= b,\,\,B_{n+1}=b,
\end{eqnarray}
and with $A_n=B_n=0$ for all other sites $n \neq n_0 , n_0+1$. Here the Eqs.~(\ref{IS-IS}) and~(\ref{IA-IA})  refer to the  in-phase and
out-of-phase compactons, respectively, while  Eqs.~(\ref{IS-IA}) and~(\ref{IA-IS}) to compactons of  mixed-symmetry type.
\begin{figure}
\hskip -.4cm
\centerline{\includegraphics[scale=0.6]{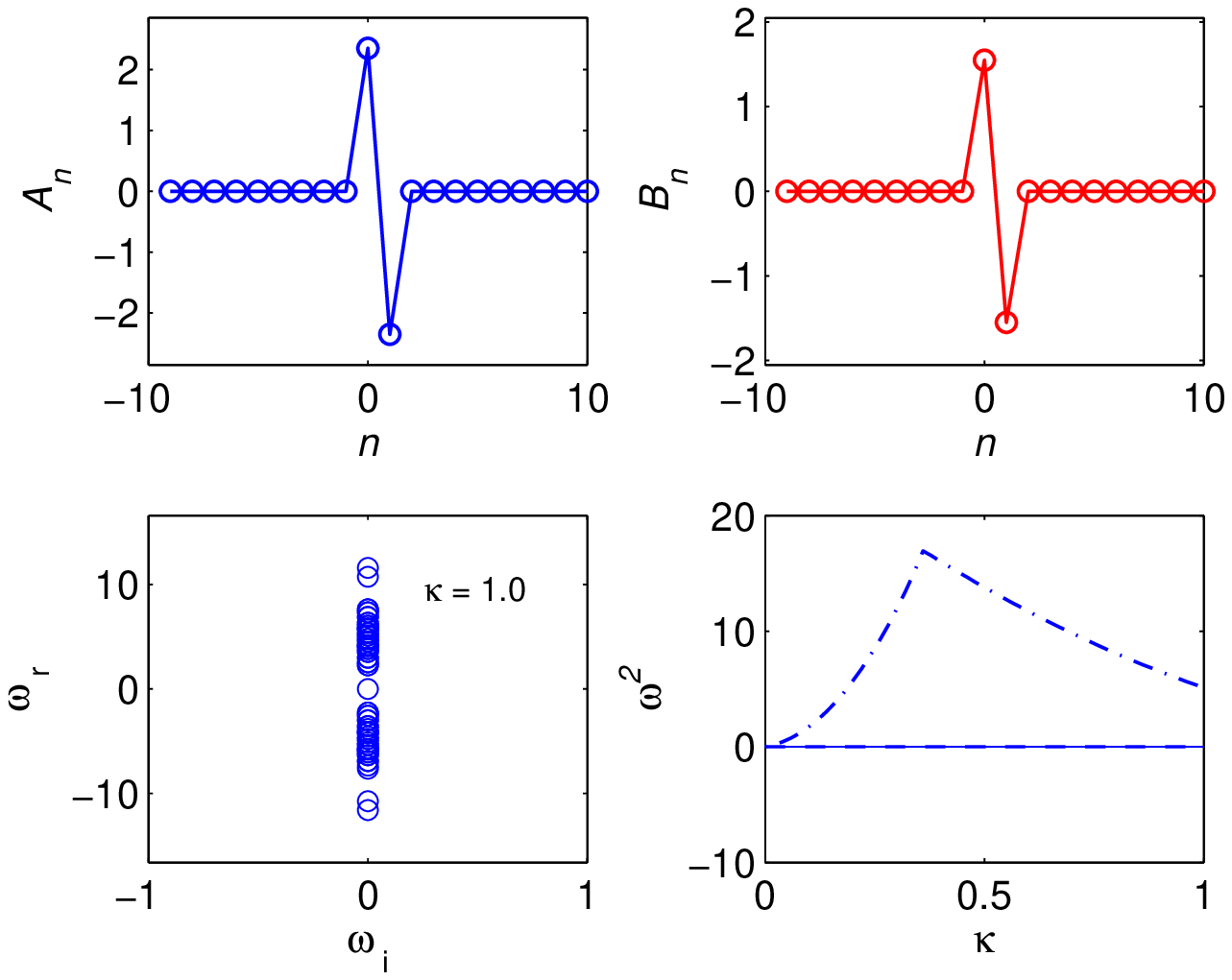}
\hskip -.7cm
\includegraphics[scale=0.6]{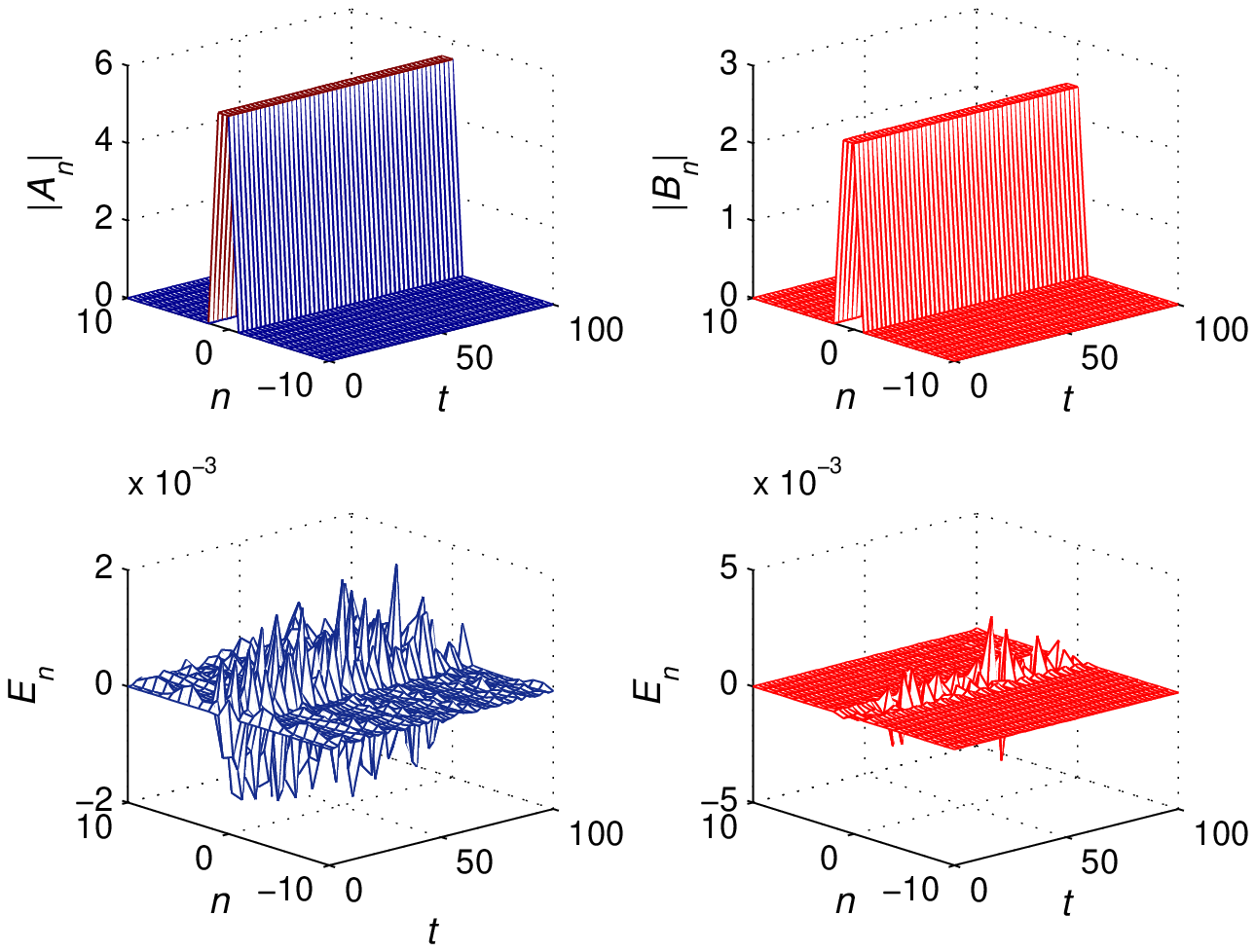}}
\caption{(Color online)
Same as in Fig.~\ref{brightbrightsingle} but for a two-site out-of-phase B-B compacton corresponding to the amplitude pattern given in Eq.~(\ref{IA-IA}) with $a, b,$ fixed as the square root of the second and first zero of the Bessel function $J_0$, respectively.}
\label{bbtwositeoutkp}
\end{figure}
By substituting the above expressions into Eqs.(\ref{mubg1},\ref{mubg2}) one obtains the same condition in Eq.~(\ref{aconst}) as before, but  with chemical potentials given by:
\begin{eqnarray}
\mu_{u} & =&-\gamma_{1}a^{2}-\gamma_{12}^{(0)}b^{2}+ \pm \kappa_{1},\nonumber \\
\mu_{v} & =&-\gamma_{2}b^{2}-\gamma_{12}^{(0)}a^{2}+ \pm \kappa_{2,}\label{inphase}
\end{eqnarray}
where the combinations of signs, $(-,-)$, $(+,+)$ in front of the terms $\kappa_{1}, \kappa_{2}$, refer to the in-phase and out-of-phase
compactons, respectively, while the other combinations $(-,+)$ and $(+,-)$, refer to mixed-symmetry compactons in (\ref{IS-IA}), (\ref{IA-IS}), respectively.

Typical result for the  in-phase and out-of-phase two-site B-B compactons are shown on Fig.~\ref{brightbrightwosite} and in  Fig.~\ref{bbtwositeoutkp}, respectively. We see that in both cases the excitation are very stable for a wide parameter range and the stability predicted by the linear analysis is corroborated by the direct numerical integrations of the original (unaveraged) system.   Similar results are obtained also for two site compactons of  mixed-symmetry type, but not shown here for brevity.

\subsection{Three-sites compactons}
\begin{figure}[!h]
\hskip -.4cm
\centerline{\includegraphics[scale=0.6]{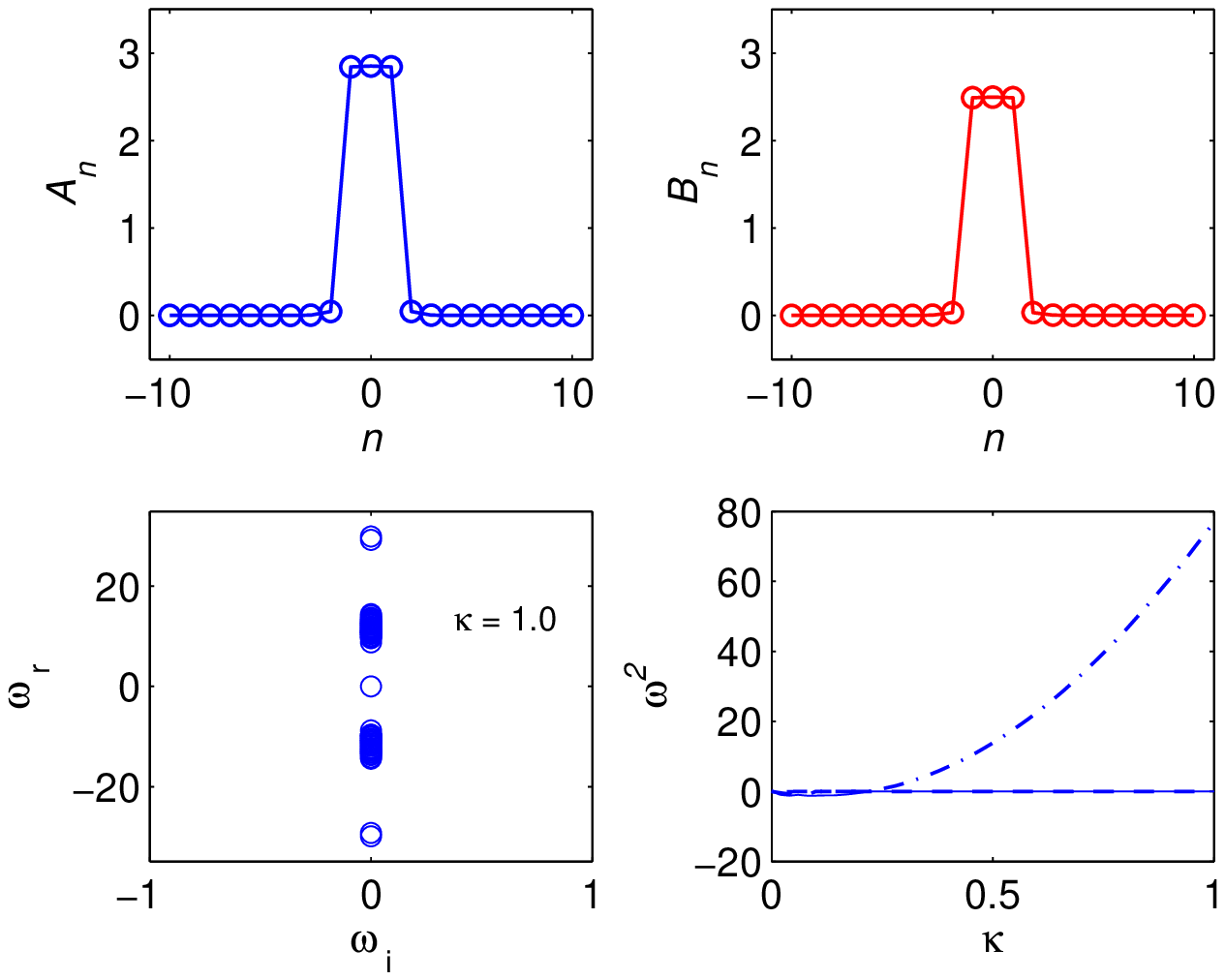}
\hskip -.8cm
\includegraphics[scale=0.6]{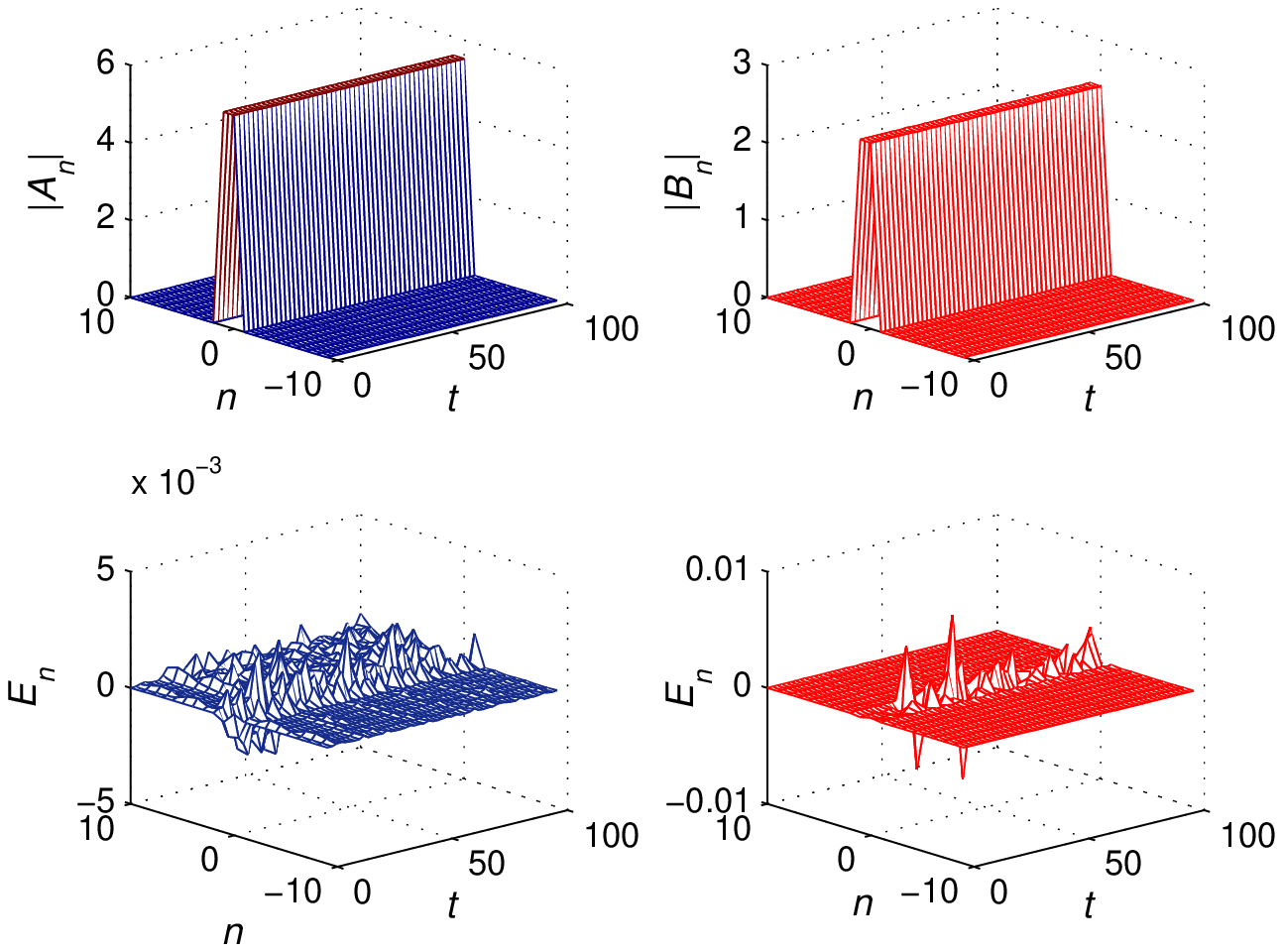}}
\caption{(Color online)
Same as in Fig.~\ref{brightbrightsingle} but for
a three-site B-B compacton corresponding to the amplitude pattern given in Eq.~(\ref{3-site-ansatz}) with $a_2, b_2,$ fixed as the square root of the third and fourth zero of the Bessel function $J_0$, respectively, and with $a_1, b_1$ obtained from  Eqs.  as: $a1=   , b_1= $. All other  parameters are  fixed as in Fig.~\ref{brightbrightsingle}.}
\label{threebbkap}
\end{figure}

Compactons of larger size can  be found in similar manner. For example, we consider here only the case of three-site B-B compactons, corresponding to the pattern
\begin{eqnarray}
&&
A_{n_{0}}=a_1,\,A_{n_{0}\pm1}=a_2,\,A_{n_{0}\pm2}=0, ..., \nonumber  \\
&&
B_{n_{0}}=b_{1},\,B_{n_{0}\pm1}=b_{2},\,B_{n_{0}\pm2}=0, ...,
\label{3-site-ansatz}
\end{eqnarray}
one finds the chemical potentials are given by
\begin{eqnarray*}
\mu_{u}= & -&\gamma_{1}a_{2}^{2}-\gamma_{12}^{(0)}b_{2}^{2}-\kappa_{1}a_{1}J_{0}\left(\xi_{v}\right)/a_{2} - 2\alpha\kappa_{2}b_{1}b_{2}J_{1}\left(\xi_{v}\right)\\
\mu_{v}= & -&\gamma_{2}b_{2}^{2}-\gamma_{12}^{(0)}a_{2}^{2}-\kappa_{2}b_{1}J_{0}\left(\xi_{u}\right)/b_{2} - 2\alpha\kappa_{1}a_{1}a_{2}J_{1}\left(\xi_{u}\right)
\end{eqnarray*}
with $\xi_{u}=\alpha\left(a_{1}^{2}-a_{2}^{2}\right)$ and $\xi_{v}=\alpha\left(b_{1}^{2}-b_{2}^{2}\right)$. The boundary amplitudes $a_2, b_2$ are given by the zero tunneling condition as in Eq.~(\ref{aconst}), e.g.
\begin{eqnarray}
J_{0}(\alpha a_2^{2})=0,\ a_2^{2}=\xi_{1}/\alpha,\nonumber \\
J_{0}(\alpha b_2^{2})=0,\ b_2^{2}=\xi_{2}/\alpha,
\label{3abconst}
\end{eqnarray}
and the central amplitudes $a_1, b_1,$ are obtained by solving the following equations:
\begin{eqnarray}
a_{1}a_{2}\left[ \gamma_{1}\left(a_{2}^{2}-a_{1}^{2}\right)+
\gamma_{12}^{(0)}\left(b_{2}^{2}-b_{1}^{2}\right)-2\alpha\kappa_{2}b_{1}b_{2}\left(2J_{1}
\left(\alpha\xi_{u}\right)+J_{1}\left(\alpha\xi_{v}\right)\right)\right]
\nonumber \\
\qquad\qquad\qquad\qquad\qquad\qquad\qquad\qquad\qquad
=\kappa_{1}\left(a_{2}^{2}-2a_{1}^{2}\right)J_{0}\left(\xi_{v}\right),\\
b_{1}b_{2}\left[ \gamma_{2}\left(b_{2}^{2}-b_{1}^{2}\right)+
\gamma_{12}^{(0)}\left(a_{2}^{2}-a_{1}^{2}\right)-2\alpha\kappa_{1}a_{1}a_{2}\left(2J_{1}
\left(\alpha\xi_{v}\right)+J_{1}\left(\alpha\xi_{u}\right)\right)\right] \nonumber \\
\qquad\qquad\qquad\qquad\qquad\qquad\qquad\qquad\qquad
=\kappa_{2}\left(b_{2}^{2}-2b_{1}^{2}\right)J_{0}\left(\xi_{u}\right).
\label{3site-eqs}
\end{eqnarray}
Typical results for three-site B-B compactons are  shown in Fig.~\ref{threebbkap}.

Before closing this section, it is worth to we note that in spite of the different averaged equations,  the existence conditions for one-site and two-site compacton are the same as for the  the intra-species SNM case  considered in Ref.~\cite{AHSU}. This fact can be easily understood by noting that under the interchange of variables $\theta_1^+ \leftrightarrow  \theta_2^+$ in the nonlinear dispersive part of the averaged Hamiltonian (\ref{Hav}), and  with  the identification of parameters $\gamma_1,\gamma_2, \gamma_{12}^{0}$ with $\gamma_1^{(0)}, \gamma_2^{(0)}, \gamma_{12}$, respectively, the Hamiltonian in Eq.~(\ref{Hav})  reduces to the one of the intra-species SNM case considered in  Ref.~\cite{AHSU} (see Eq.~(14) of Ref.~\cite{AHSU}). Clearly, this transformation is invariant for one-site compactons because, due to  the vanishing of the amplitudes at all sites $n\neq n_0$ (all the matter being localized on the site $n_0$), the nonlinear dispersive terms are automatically zero. This explain why the existence conditions for the single-site compactons are the same for inter- and intra-species SNM. A similar situation occurs also for two-site compactons because, due to the particular symmetry of the solutions, we have   $\theta_1^+ = \theta_2^+ = 0 $ and  the above transformation is  invariant for two-site compactons. In all other cases, however, e.g. for all compactons localized on more than two sites, the invariance is broken and the existence conditions are  different for inter- and  intra-species SNM cases. It should also be remarked  here that since the  averaged equations of motion for inter- and intra-species SNM cases are intrinsically different, stability properties of binary compactons  will always be different (including one-site and two-site compactons) in the two SNM cases.
\begin{figure}
\hskip -.3cm
\centerline{\includegraphics[scale=0.6]{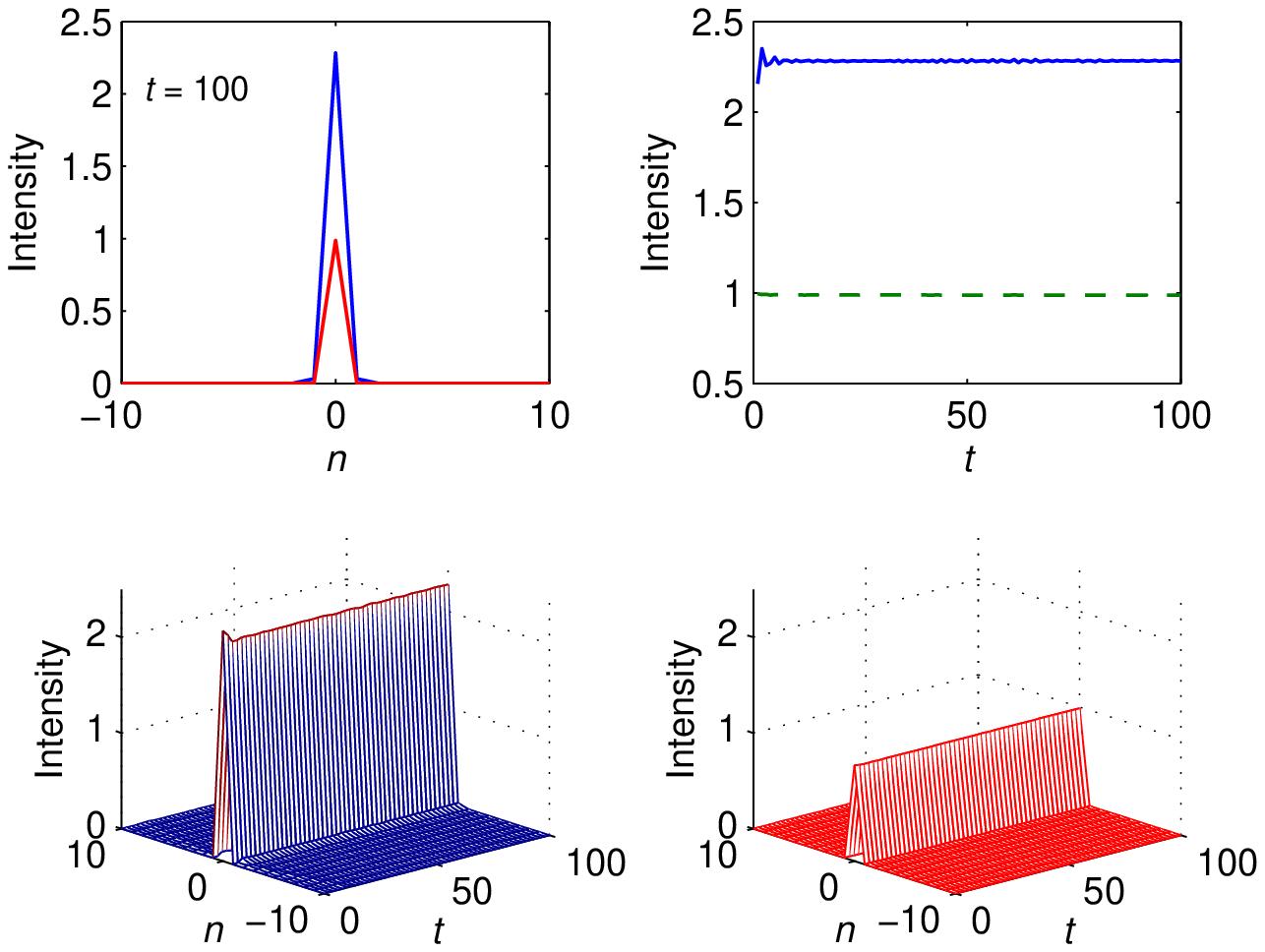}
\hskip -.8cm
\includegraphics[scale=0.6]{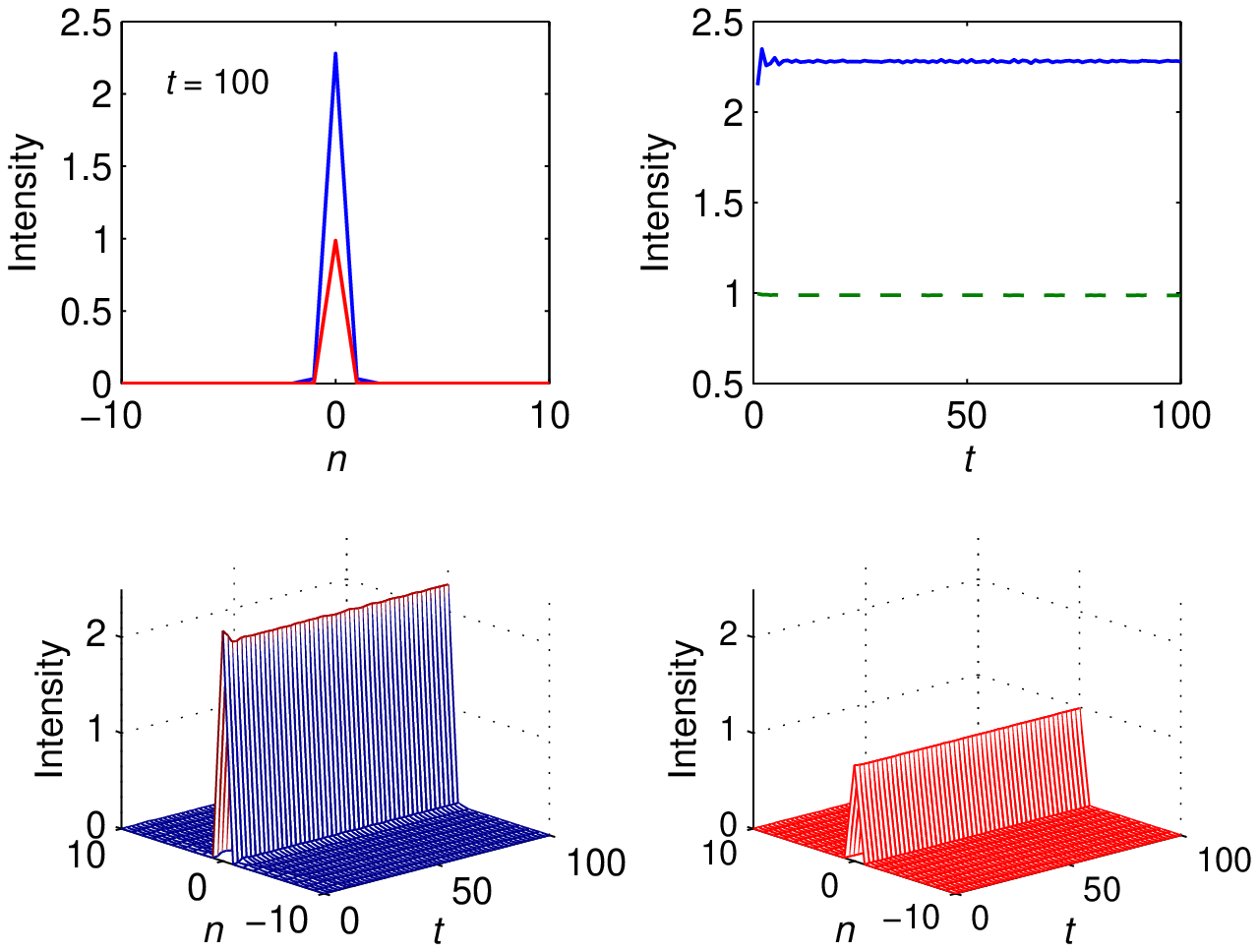}}
\caption{(Color online) Top panels. First (upper curve) and second (lower curve) component profiles of a single site B-B quasi-compacton  at time $t=100$ (first and third top panels from the left), and density time evolution  at the middle site $n_0=0$ (second and fourth panel from the left), as obtained
from numerical integrations of Eqs. (\ref{eq.av1},\ref{eq.av2}) (first and second panels from the left) and of Eq.~(\ref{vdnlse}) (third and fourth panels from the left), respectively.
Bottom panels shown the full time evolution of the quasi-compacton first  and second components obtained from Eqs. (\ref{eq.av1},\ref{eq.av2}) (first and second panels from the left) and from Eq.~(\ref{vdnlse}) (third and fourth panels from the left), respectively.
In all panels the initial conditions were taken as $\left|A_{0}\right|^{2}=2.4048$ (first root of Bessel function), $\left|B_{0}\right|^{2}=1.0$ while the parameter values were fixed as  $\gamma_{1}=\gamma_{2}=1,\gamma_{12}^{(0)}=1.0, \kappa_{1}=\kappa_{2}\equiv\kappa=0.5\,\textrm{and}\,\alpha_{1}=\alpha_{2}=1$. The numerical  integration
of Eq.~(\ref{vdnlse}) is made with the same  modulation function $\gamma_{12}(t)$  as in Fig.~\ref{brightbrightsingle} but with $\epsilon=0.005$.}
\label{BBg12-1}
\end{figure}

\section{Stationary quasi-compactons under inter-species SNM}
From the above analysis it follows that a necessary condition for exact stationary compactons  to exist is that the BEC densities (multiplied by $\alpha$) at the edges sites coincide with zeros of the Bessel function $J_0$. This is obviously necessary for the zero tunneling condition in Eq.~(\ref{aconst}) to be satisfied and therefore for the  suppression of any tail in the solution. In the case  of a single site compacton this implies that  the numbers of atoms in the two components cannot be arbitrary but must be related by
\begin{equation}
\frac{N_{1}}{N_{2}}=\frac{\xi_{1}}{\xi_{2}}
\end{equation}
with $\xi_i$ two zeros (not necessarily equal) of $J_0$.
From this it also follows that  for an exact binary compacton with a given number of particles $N_{1}$ in the
first component, controlled by the parameter $\alpha$, only  a discrete set of values for $N_{2}$ are permitted.

On the other hand, it is also interesting  to look for stationary quasi-compactons in which one of the two components amplitude (say the first) is associated to a zero of the Bessel function $J_0$, while the other is not. In the following we report  results obtained  by direct numerical integrations of both the averaged Eqs. (\ref{eq.av1},\ref{eq.av2}) and the original system Eq.~(\ref{vdnlse}), for specific choices  of the  (inexact)  component and for different parameter values, with the interspecies interaction  $\gamma_{12}^{(0)}$ typically varied in the range $0 \div 1.0$ (similar results are obtained for other choices of parameters). Notice that for the above  setting, while the first component is initially taken as an exact compacton  in the uncoupled limit $\gamma_{12}^{(0)}=0$, the second component is inexact and therefore can not survive in absence of coupling (see Ref.~\cite{AKS} for the single component case).
\begin{figure}
\hskip -.3cm
\centerline{\includegraphics[scale=0.6]{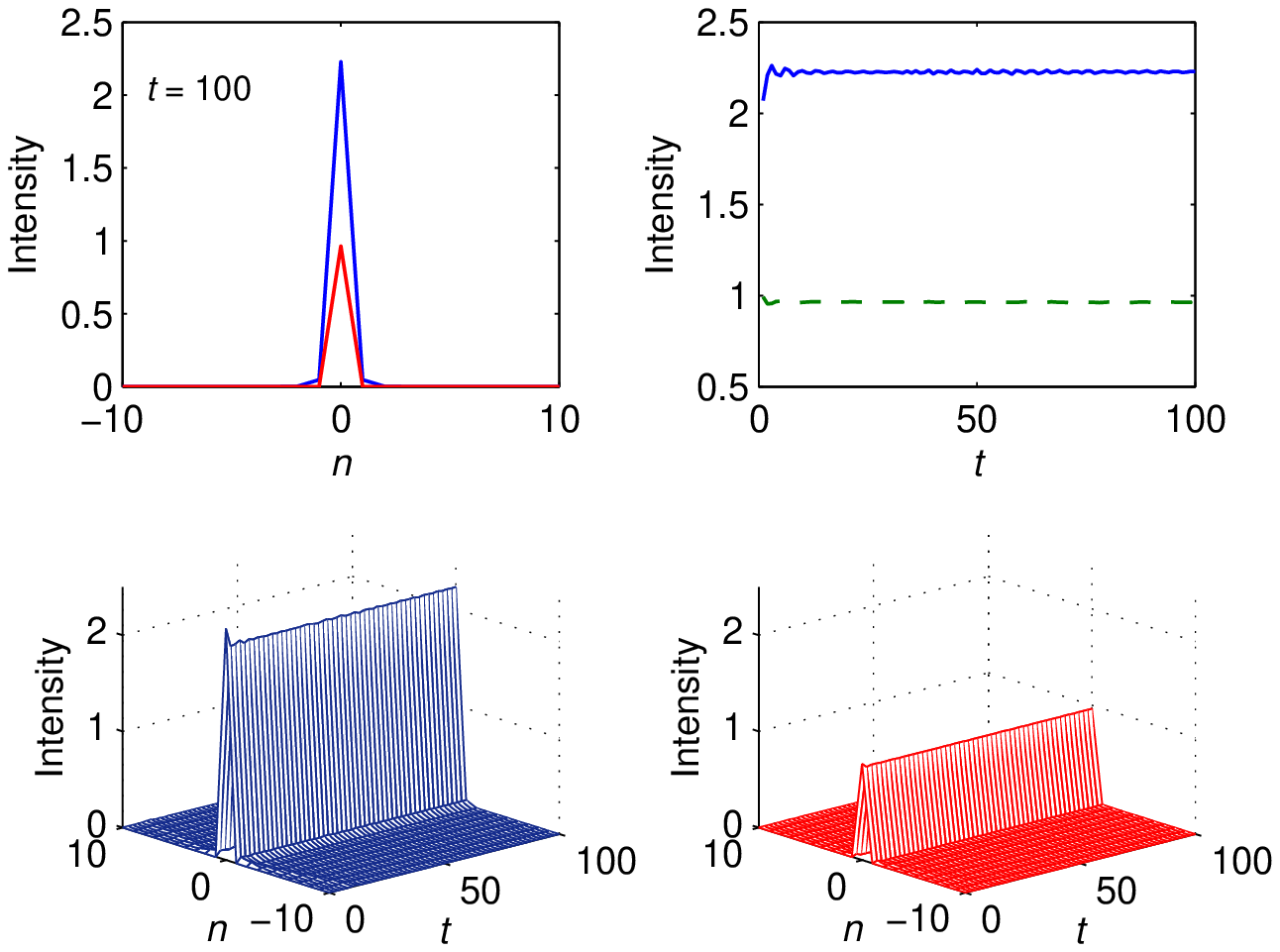}
\hskip -.8cm
\includegraphics[scale=0.6]{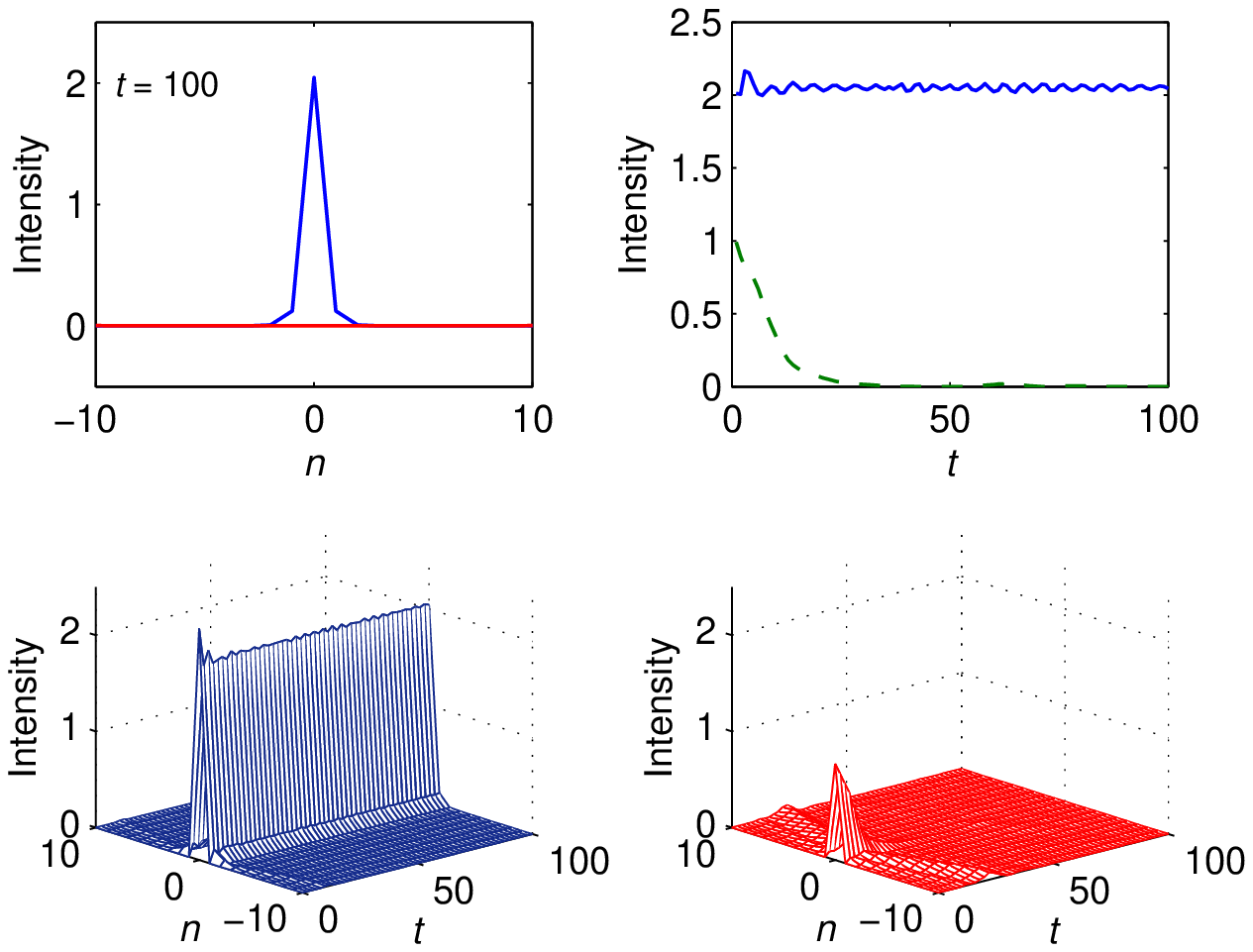}}\protect
\protect\protect\protect\caption{(Color online) First and second rows, repeating the same parameters
as in Fig. \ref{BBg12-1} except
$\gamma_{12}=0.3$ and similarly for the third and the fourth rows
but $\gamma_{12}=0.2$. The results are obtained by solving the original
Eq.~(\ref{vdnlse}). }
\label{twocoloc1s1}
\end{figure}
\begin{figure}
\hskip 1.5cm
\centerline{\includegraphics[scale=0.28]{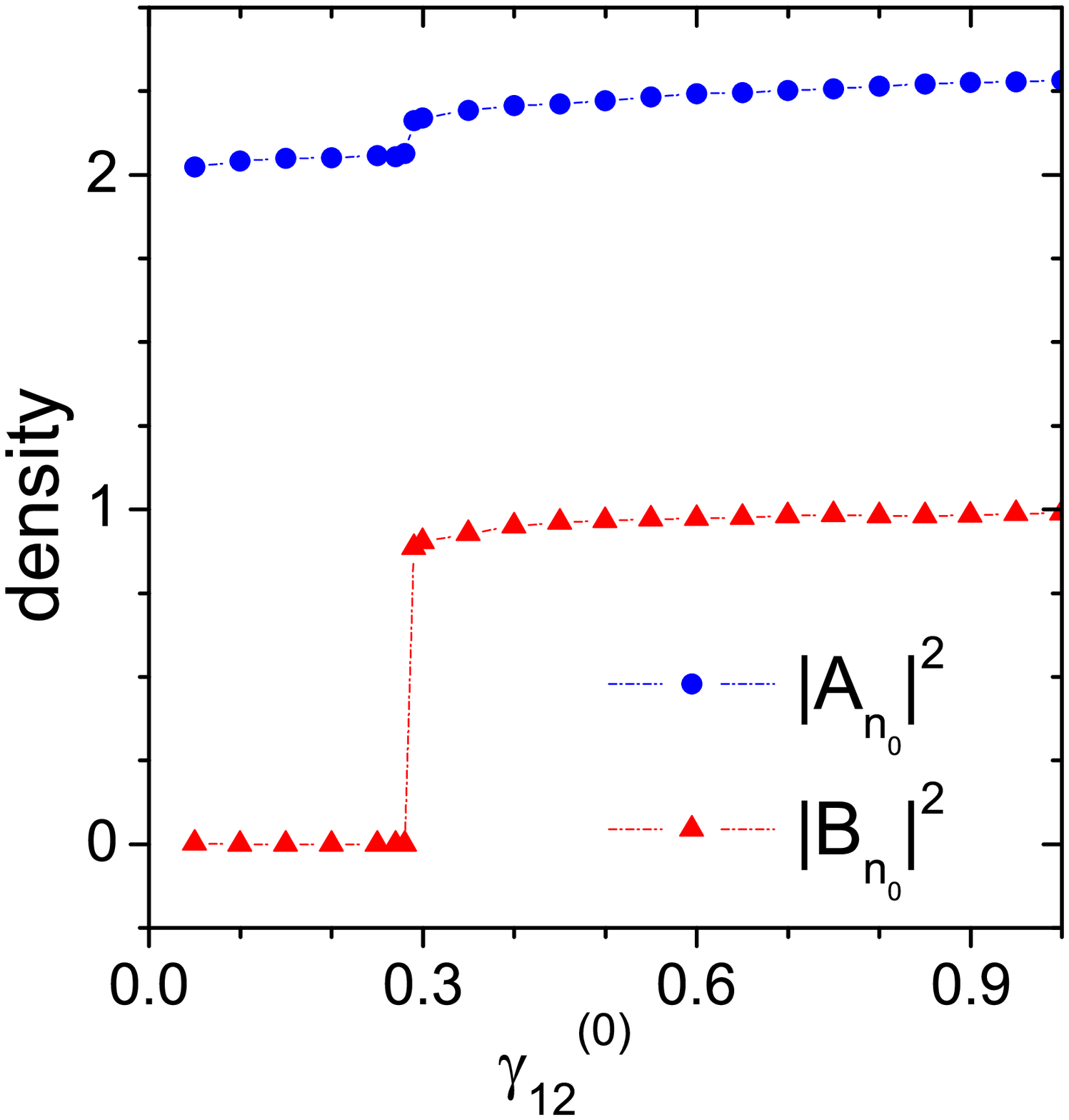}
\hskip -2cm
\includegraphics[scale=0.28]{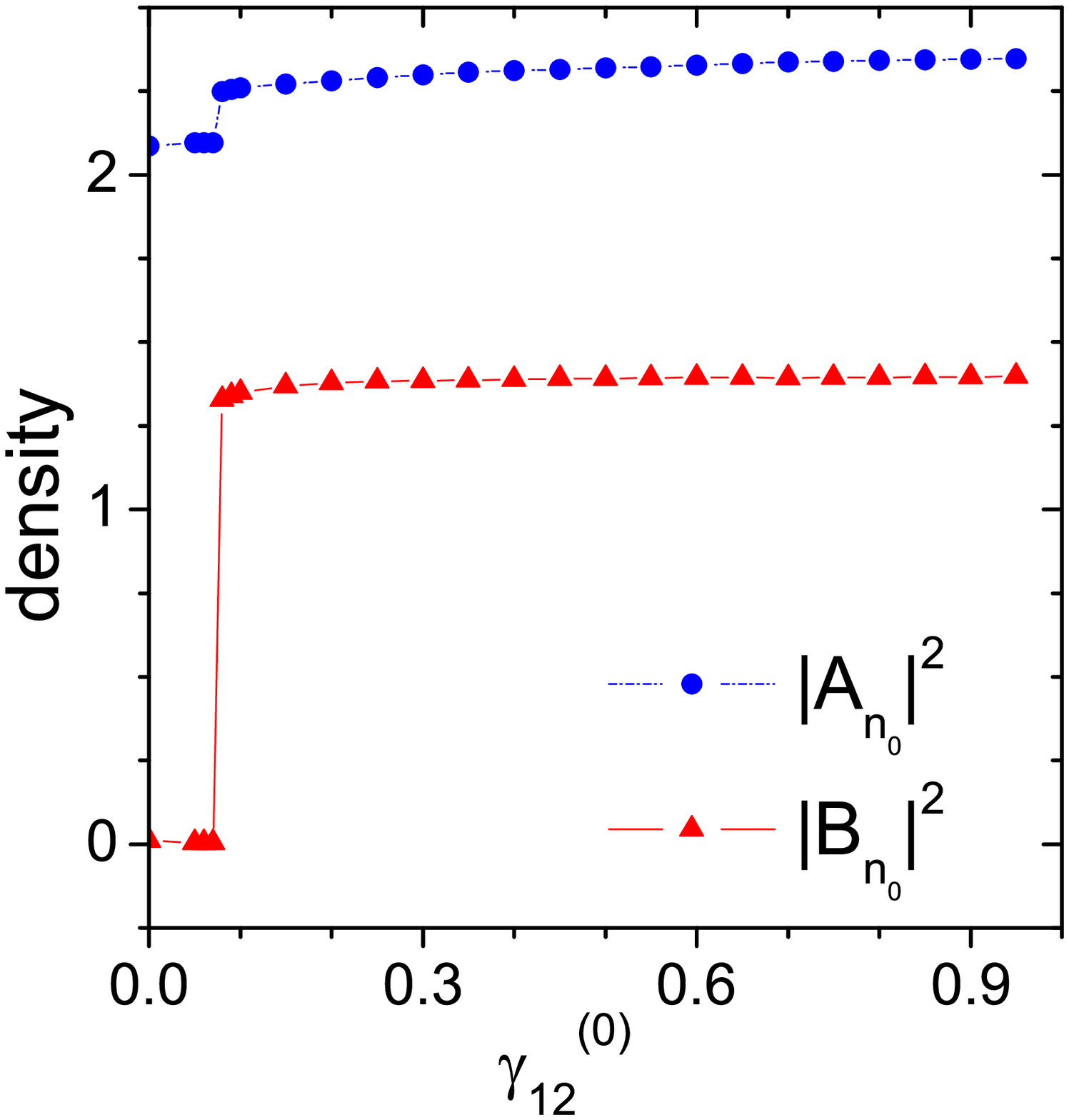}
}
\vskip -.5cm
\hskip 1.5cm
\centerline{\includegraphics[scale=0.28]{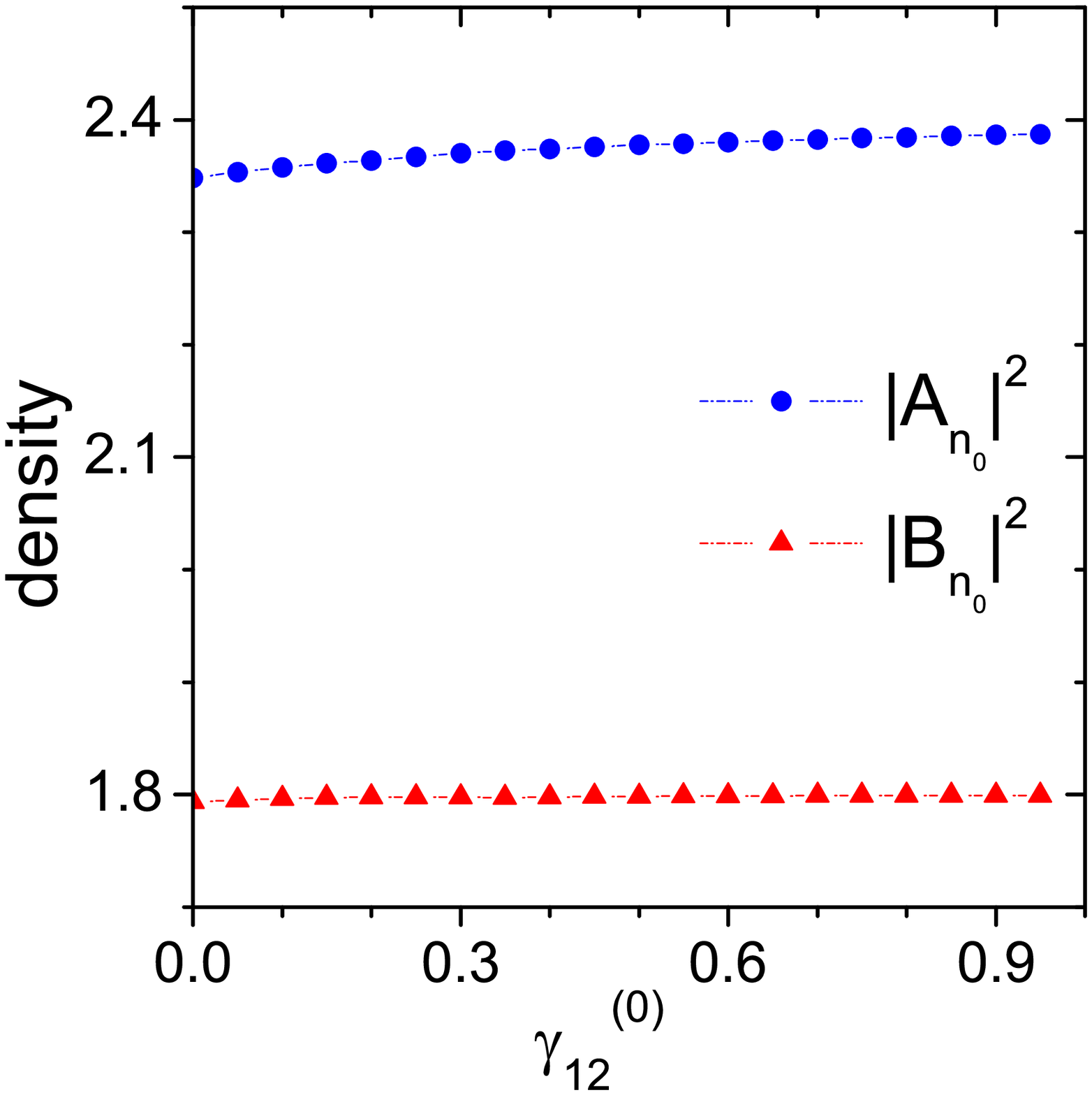}
\hskip -2cm
\includegraphics[scale=0.28]{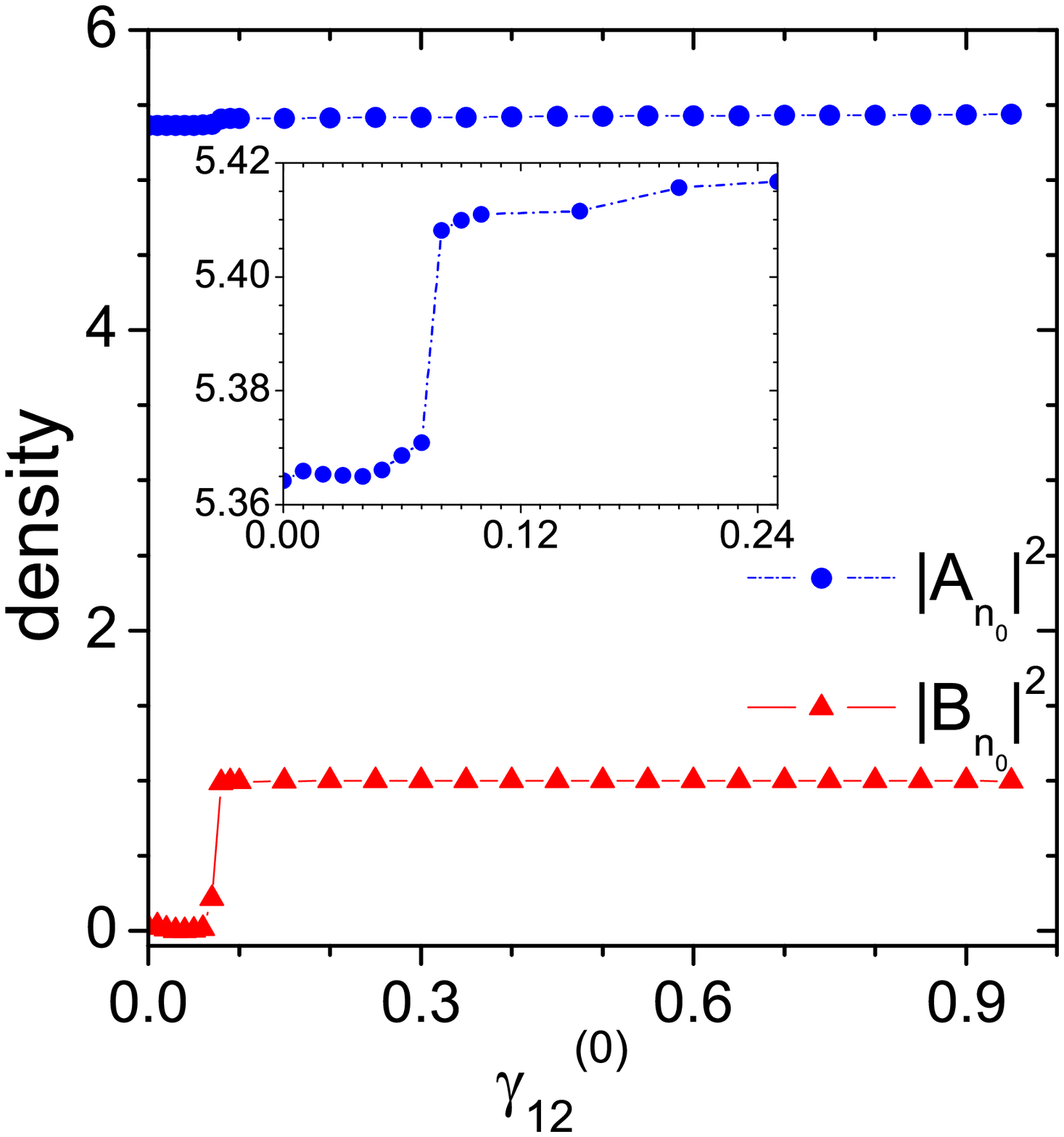}
}
\hskip -3cm
\caption{(Color online) Dependence on the parameter  $\gamma_{12}^{(0)}$ of the first and second  component  densities $|A|^2, |B|^2$, respectively, at  site $n_0$ taken at time $t=300$, as obtained from direct numerical integrations of Eq.~(\ref{vdnlse}). In the top panels and in the bottom left panel the initial condition for the first component  at site $n_0$ was taken as the first root ($\left|A_{0}\right|^{2}=2.4048$) of the $J_0$ Bessel function and zero elsewhere,  while second components at site $n_0$ were fixed as  $\left|B_{0}\right|^{2}=1.0$ (top left panel), $\left|B_{0}\right|^{2}=1.4$ (top right panel), $\left|B_{0}\right|^{2}=1.8$ (bottom left panel) and zero elsewhere. In the bottom right panel the initial condition was $\left|A_{0}\right|^{2}=5.52007$ (second zero of $J_0$) and $\left|B_{0}\right|^{2}=1.0$. The inset in this panel shows an enlargement of the delocalizing transition occurring in the first component near $\gamma_{12}^{(0)} \approx 0.075$ of the top curve. All symbols correspond to points computed numerically, joining lines being drawn just to guide eyes. Other parameters are fixed  as in Fig.~\ref{BBg12-1}. To accelerate convergence to stationary quasi-compacton at $t=300$, the dissipative function in (\ref{bdiss}) with $\eta=5$ was used during the time evolution.}
\label{Addition1}
\end{figure}

In the top panels of Figs.~\ref{BBg12-1}, ~\ref{twocoloc1s1},  are reported the profiles of the quasi-compacton components at the time $t=100$ (first and third top panels from the left)  together with their peak density at the middle sites time evolution (second and fourth top   panels from the left), for the case $\gamma_{12}=1.0$ in Fig.~\ref{BBg12-1} and for the cases $\gamma_{12}=0.3$ (first and second panels from the left) and $\gamma_{12}=0.2$ (first and second panels from the right) in  Fig.~\ref{twocoloc1s1}. As one can see, while in the first two cases the density at the middle site of the first component after  an initial drop (due to matter emission)  adjusts to a new practically constant value with small oscillations around it, and  the density of the second component remains practically constant in time, in the last case (e.g. $\gamma_{12}=0.2$) the second component cannot survive and quickly decays into background radiation.  This ia also clearly seen from the corresponding bottom panels showing the space-time evolution of the two BEC components. Actually this decay is associated to the occurrence of  a delocalizing transition (see below) similar to the ones observed for single component BEC~\cite{Baizakov} and binary BEC mixtures in deep OL~\cite{cruz2009}. We also remark that in Fig.~\ref{BBg12-1} the time evolutions are computed both with the averaged equations (first and second panels from the left)   and with the original (unaveraged) system (first and second panels from the right), obtaining  practically identical  results. This is true for all numerically investigated cases  provided the parameter $\epsilon$ is small enough (typically $\epsilon < 0.01$) for the SNM limit to be valid. In the following  we remain strictly in this limit and  report the time evolution only for one (averaged or original) system .

\begin{figure}
\hskip 1.5cm
\centerline{\includegraphics[scale=0.28]{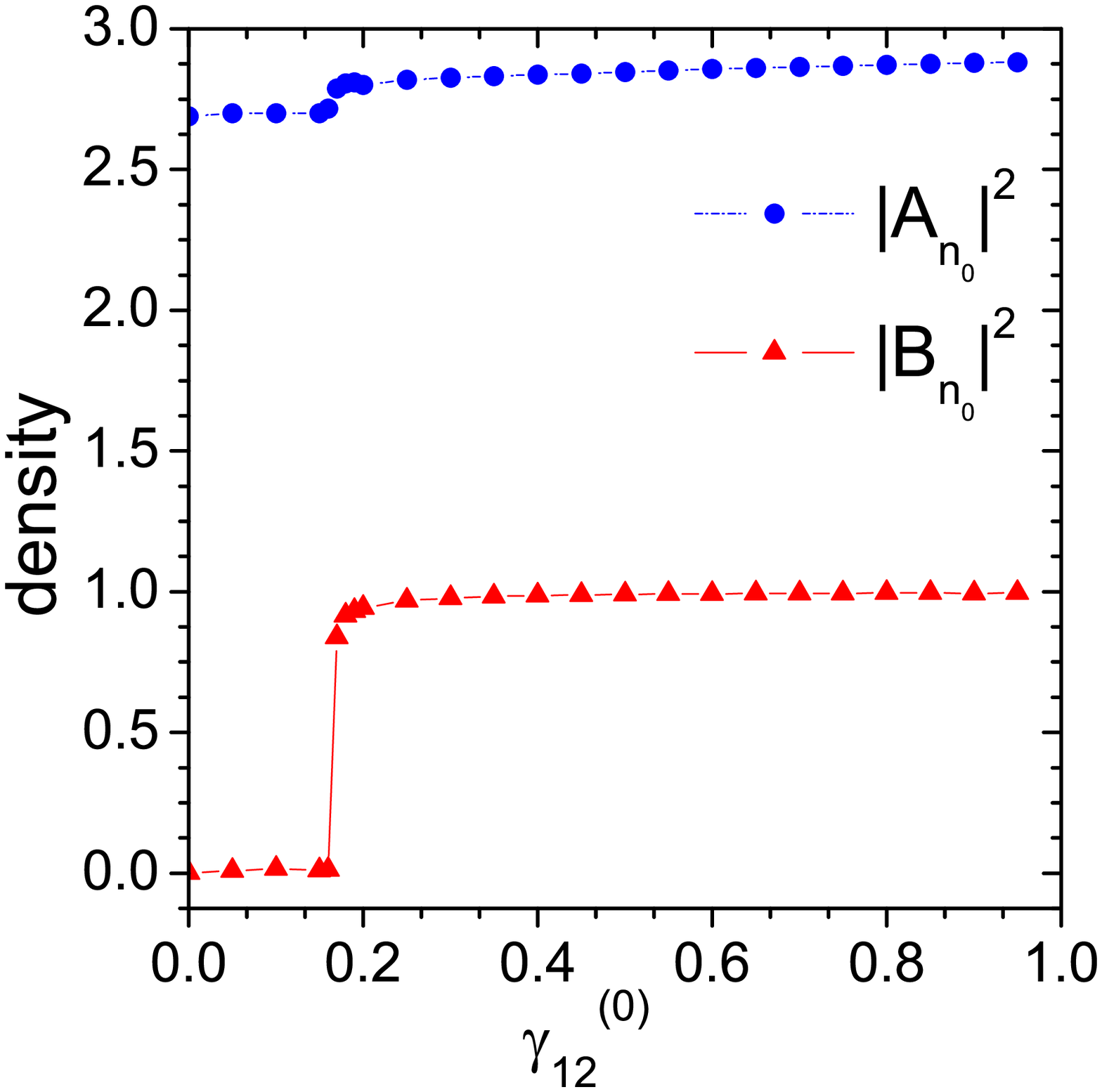}
\hskip -2cm
\includegraphics[scale=0.28]{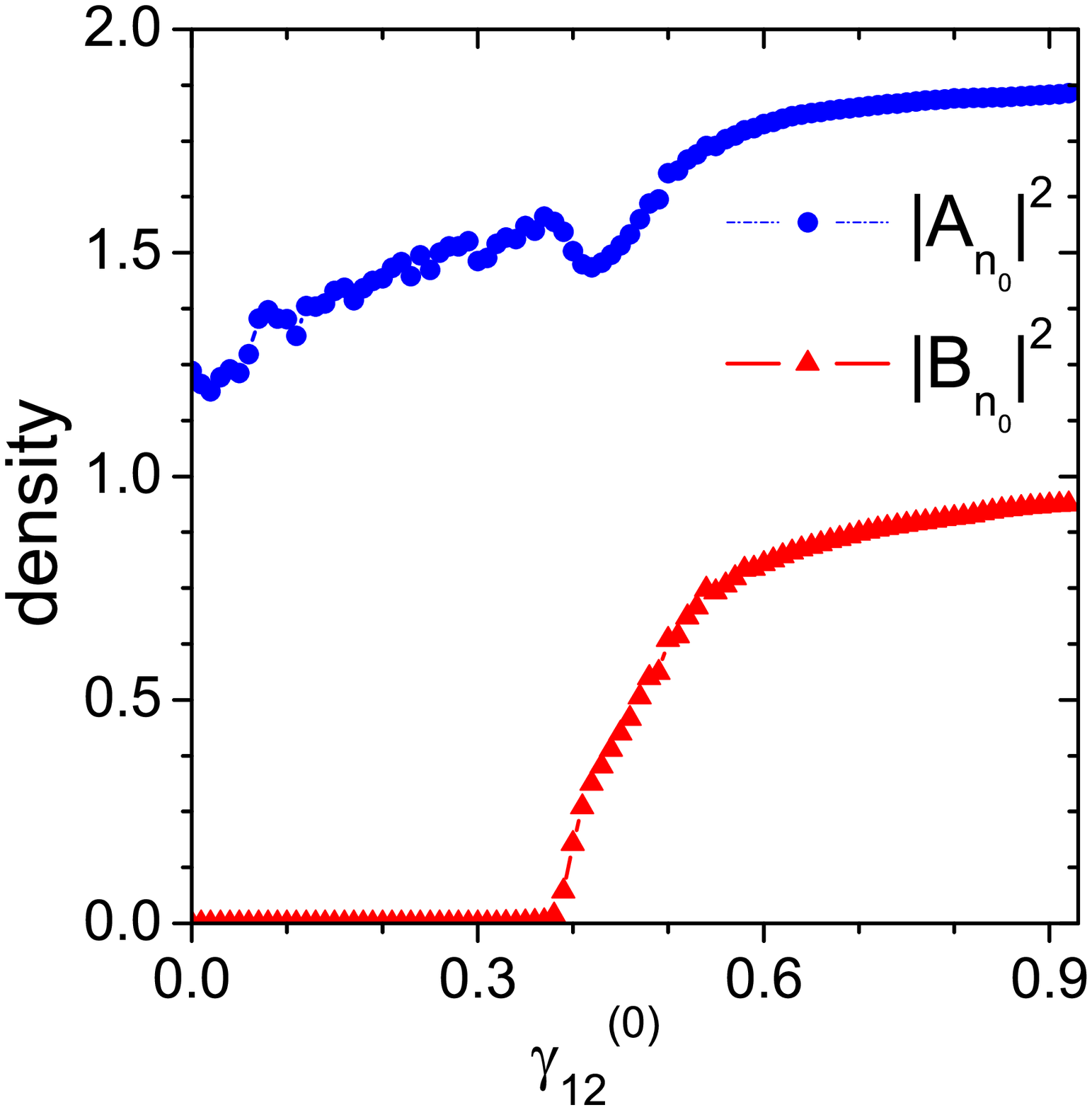}
}
\hskip -3cm
\caption{(Color online) Same as in Fig.~\ref{Addition1} but for the case in which both components deviate from the exact zero tunneling condition. In the left and right panels the first component at central site $n_0$ was initially fixed as initial conditions at the central site were fixed as
$\left|A_{0}\right|^{2}=3.0$ and $\left|A_{0}\right|^{2}=2.0$, respectively, and zero elsewhere, while the second component at site $n_0$ was fixed as $\left|B_{0}\right|^{2}=1.0$ for both panels, and zero elsewhere. All other parameters are the same as in Fig.~\ref{BBg12-1}.
}
\label{Addition1bis}
\end{figure}

From Figs.~\ref{BBg12-1},~\ref{twocoloc1s1}, we see that for fixed system parameters and initial conditions  the existence  of stable quasi-compactons crucially depends on the mean inter-species interaction parameter $\gamma_{12}^{(0)}$. More detailed  investigations show that stable quasi-compactons  exist in the SNM limit under very generic initial conditions (see below) provided the strength of the mean interspecies scattering length is above a critical threshold. Moreover, this threshold disappears  when the initial amplitudes are relatively close to  zeros of $J_0$.  This can be  seen from  Figs.~\ref{Addition1},~\ref{Addition1bis},  where the dependence of the BEC densities at the central site $n_0$  on the parameter  $\gamma_{12}^{(0)}$ is reported. Calculations here were done by numerically solving the the averaged equations of motion   for different initial conditions, scanning $\gamma_{12}^{(0)}$ in the interval 0,1 with typical increments of $0.1$ (reduced to  $0.01$ close to the delocalizing transition) and by introducing  dissipative boundary conditions to eliminate the emitted radiation from the system and accelerate  convergence to stationary quasi-compactons. This  was achieved by adding dissipative terms of the form
$i\xi_{n}u_{n},\, i\xi_{n}v_{n}$ (resp. $i \xi_{n}U_{n},\, i \xi_{n}V_{n}$) into the original (res. the averaged) equations for first and second component, respectively, with the function $\xi_{n}$ taken as
\begin{equation}
\xi_{n}=\eta \textrm{ sech}^{2}\left(n-n_{b}\right),
\label{bdiss}
\end{equation}
with  $n_{b}$ denoting the boundary sites and the real parameter $\eta$ controlling the strength of the dissipation. In all the panels  of Fig.~\ref{Addition1} the initial conditions for the first component were taken to satisfy the exact compacton zero tunneling condition in correspondence of the first (top panels and bottom left panel)  and second (bottom right panel)
zero of the  Bessel function,  while the second component was fixed as $\left|B_{0}\right|^{2}=1.0$ for top left and bottom right panels, and as $\left|B_{0}\right|^{2}=1.4$ and $\left|B_{0}\right|^{2}=1.8$ for the top right and bottom left panels, respectively. From  the top left panel of Fig.~\ref{Addition1} we see that a sharp transition (existence threshold) occurs at $\gamma_{12}^{(0)} \approx 0.29$. As the deviation of the inexact second component from the zero tunneling condition is reduced, keeping fixed all other parameters,  the existence threshold moves toward a smaller value (see top right panel) and completely disappears  when it is further reduced (see bottom left panel). For the considered parameters we find   that the threshold disappears as the initial density of the second component  at site $n_0$ become smaller than the critical value $|B_{cr}|^2\approx 1.535$.
Similar behaviors is observed also when the exact (first) component matches the zero tunneling condition at  higher zeros of the Bessel function, as one can see from the bottom right panel of Fig.~\ref{Addition1}. In this case, however, note that the transition appears less sharp and also a bit more noisy in the tail, this being a  consequence of the larger first component amplitude involved (see enlargement in the inset) and possibly also of the longer time required to converge to the solution.
\begin{figure}
\hskip .8 cm
\includegraphics[scale=0.33]{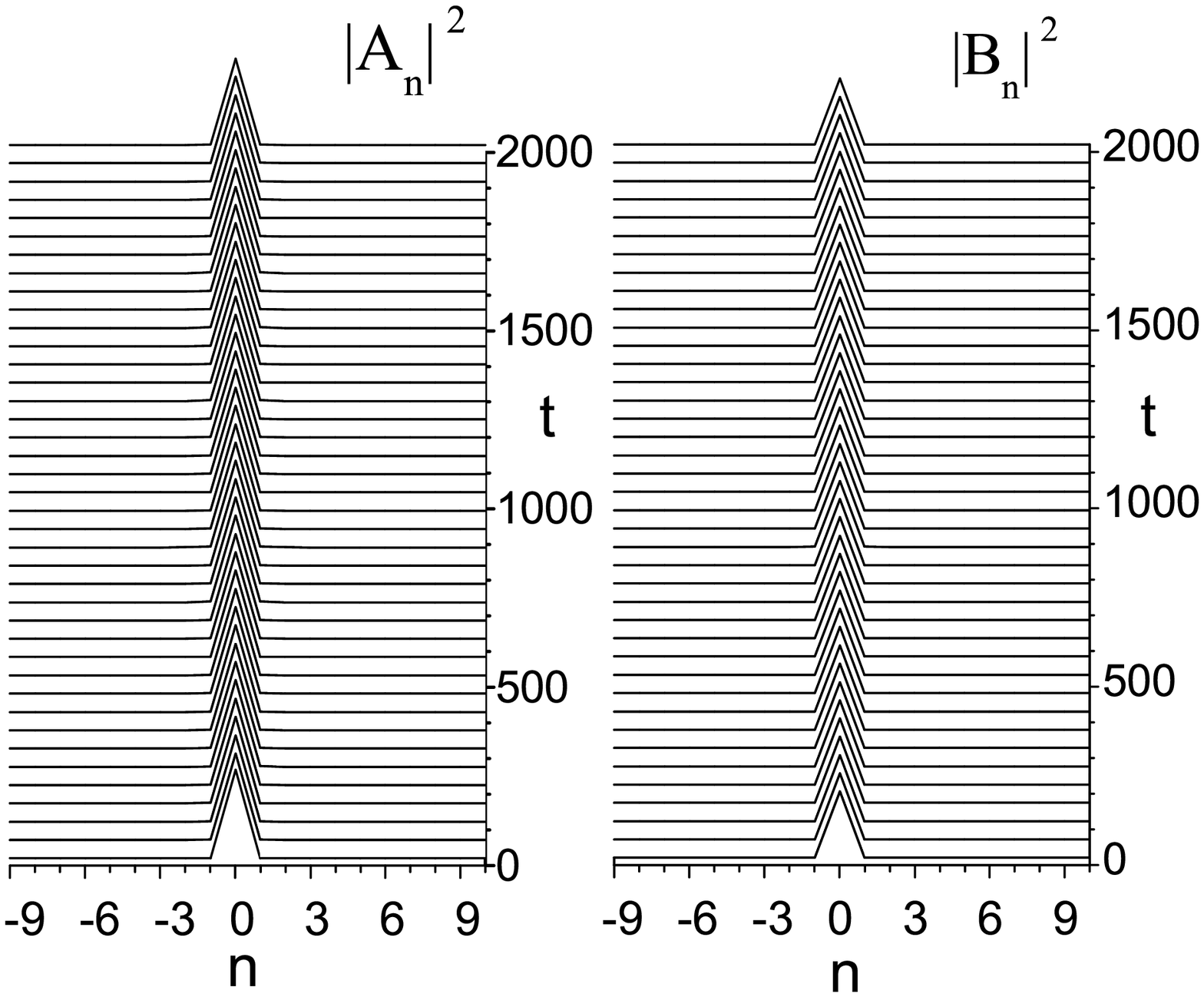}
\hskip -.0cm
\includegraphics[scale=0.33]{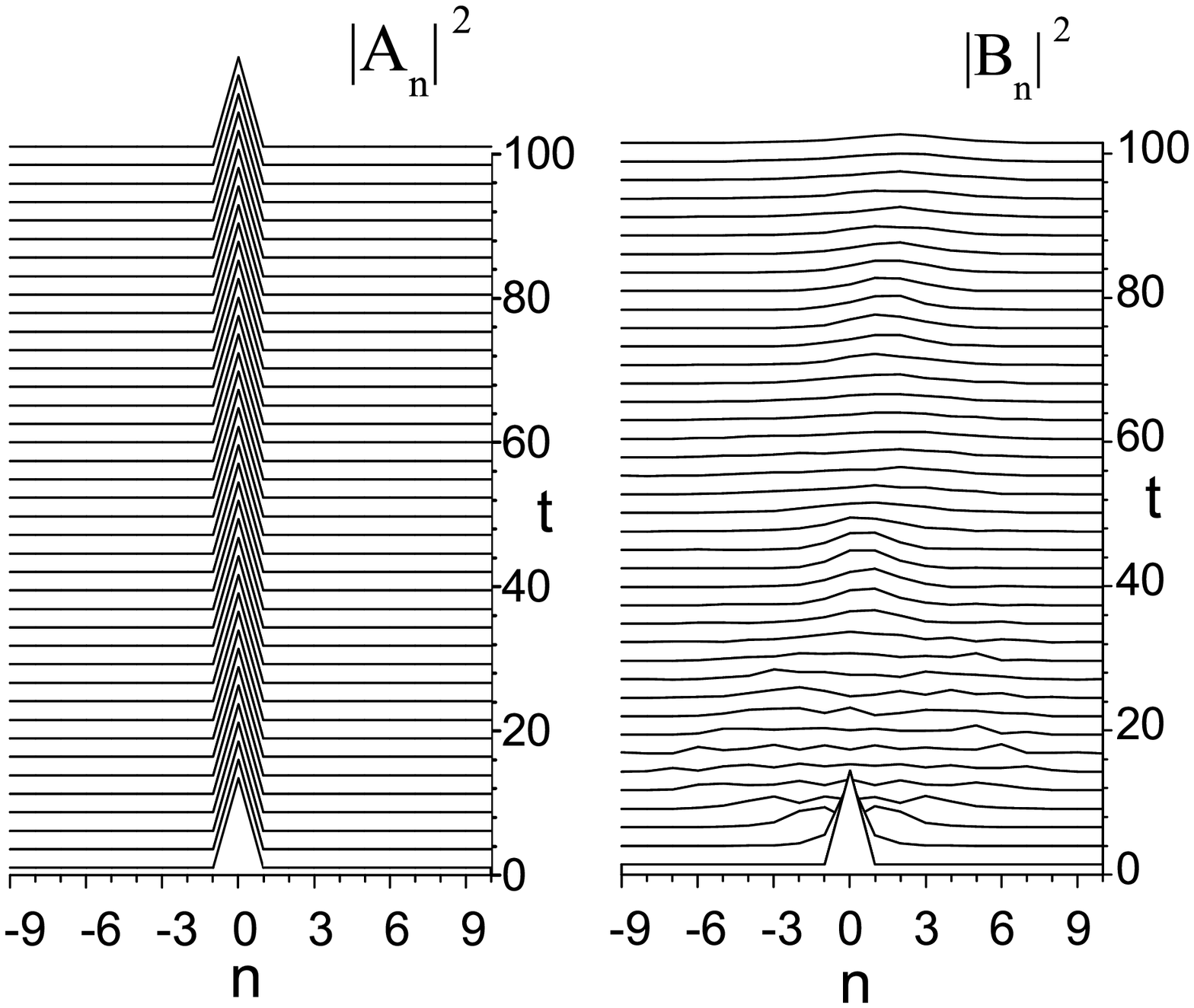}
\vskip -3.5cm
\caption{(Color online) Time evolution of a one-site quasi-compacton (first and second panel from the left) obtained from numerical integration of Eq.~(\ref{vdnlse}) with $\gamma_{12}^{(0)}=0$ and initial conditions $\left|A_{0}\right|^{2}=2.4048$ (first root of Bessel function), $\left|B_{0}\right|^{2}=1.8$.
 Other parameters as  fixed as for the  bottom left panel Fig.~\ref{BBg12-1}. The right two panels show the corresponding  time evolution for the  intra-species SNM case, (Eqs. (12), (13) of Ref.~\cite{AHSU}), with the same initial condition and parameters values as in  the left two panels and with  $\gamma_1^{(0)} \equiv \gamma_1=1.0, \gamma_2^{(0)} \equiv \gamma_2=1.0, \gamma_{12} \equiv \gamma_{12}^{(0)}=0$.
}
\label{Addition3}
\end{figure}

From  Fig.~\ref{Addition1} it is also interesting to note  that for relatively small initial deviations from exact zero tunneling conditions there are no delocalizing transitions and it is possible to have quasi-compactons even when  the inter-species mean scattering length is detuned to zero: $\gamma_{12}^{(0)}=0$. In this case it is  interesting to compare the quasi-compacton  time evolution obtained from the inter- and  intra-species SNM, using the same initial conditions and parameter values for the two cases.  This is done in Fig.~\ref{Addition3} from which we see that the  lacking of nonlinear dispersion coupling makes impossible to sustain the quasi-compacton in the intra-species SNM case.

Finally,  in Fig.~\ref{Addition1bis} we investigate the case in which  both components deviate from the exact zero tunneling conditions. As one can see, quasi-compactons exist also in this case but  existence thresholds and sharpness of the delocalizing transition depend much more on how the  deviations of the initial conditions from the exact zero tunneling condition are taken.
In the left panel of the figure it is shown the case in which the first component density is above  the  first zero of the Bessel function and  the second component is below (e.g. $\left|A_{0}\right|^{2}=3.0,\;\left|B_{0}\right|^{2}=1.0$), while in the right panel the case in which  both components smaller than the first zero of the Bessel function (e.g. $\left|A_{0}\right|^{2}=2.0,\;\left|B_{0}\right|^{2}=1.0$). We see that in the first case the transition is sharp and similar to the ones shown in Fig.~\ref{Addition1}, while in the second case the transition is not sharp and very noisy with  both components dropping from  their initial values much more (due to a larger emitted background radiation adsorbed, in our simulation, at the ends of the chain). Quite remarkably, also in this case quasi-compacton  survive on a very long time scale similarly to the binary compacton shown in the first two left panels of Fig.~\ref{Addition1bis}.

\section{Discussion and Conclusions}

Before closing this paper let us discuss conditions for possible experimental
observation of the above results. Example of boson-boson mixtures which can be considered are $^{39}$K-$^{87}$Rb, $^{41}$K-$^{87}$Rb, $^{85}$Rb-$^{87}$Rb.
At the present binary BEC mixtures are observed for the system  $^{85}$Rb-$^{87}$Rb~\cite{Rapp} and  $^{41}K-^{87}Rb$ ~\cite{Inguscio}. The inter-species scattering length can
be varied in time by means of the Feshbach resonance technics.

Possible experimental implementations of the results discussed in this paper  could be made
by considering $^{85}Rb-^{87}Rb$ boson-boson mixtures loaded in the deep optical lattices~\cite{Rapp,Thalhammer,Catani}. The inter-species scattering length can be varied in time by means of the Feshbach resonance technics (see the recent review~\cite{REV-MOD-PHYS} and references therein). In this case the variation the inter-species scattering length $a_{12}$
can be described by the formula
\begin{equation}
a_{12}=a_{bg}(1-\frac{\Delta}{B-B_{R}}),
\label{FR}
\end{equation}
with $a_{bg}$  denoting the background scattering length, e.g. the one far from the resonance, $B_{R}$ the resonance position where the scattering length diverges, and  $\Delta$ the resonance width~ \cite{REV-MOD-PHYS}.
By assuming an external  time dependent magnetic field of the form: $B(t)= B_R +\Delta + \delta + B_1 cos(\omega t)$, with $B_1$ a constant amplitude and  $\delta$ denoting the detuning from the zero crossing point of the scattering length, occurring at $B_R+\Delta$. Introducing the parameters $\alpha=\delta/\Delta$ and $\beta=B_1/\Delta$,  we have
\begin{equation}
\frac{a_{12}}{a_{bg}}=\frac{\alpha + \beta \cos(\omega t)}{1 + \alpha + \beta \cos(\omega t)} = a_0 + a_1\cos(\omega t)+ a_2\cos(2\omega t) +...
\label{Fourier}
\end{equation}
\begin{figure}
\centerline{\includegraphics[scale=1.]{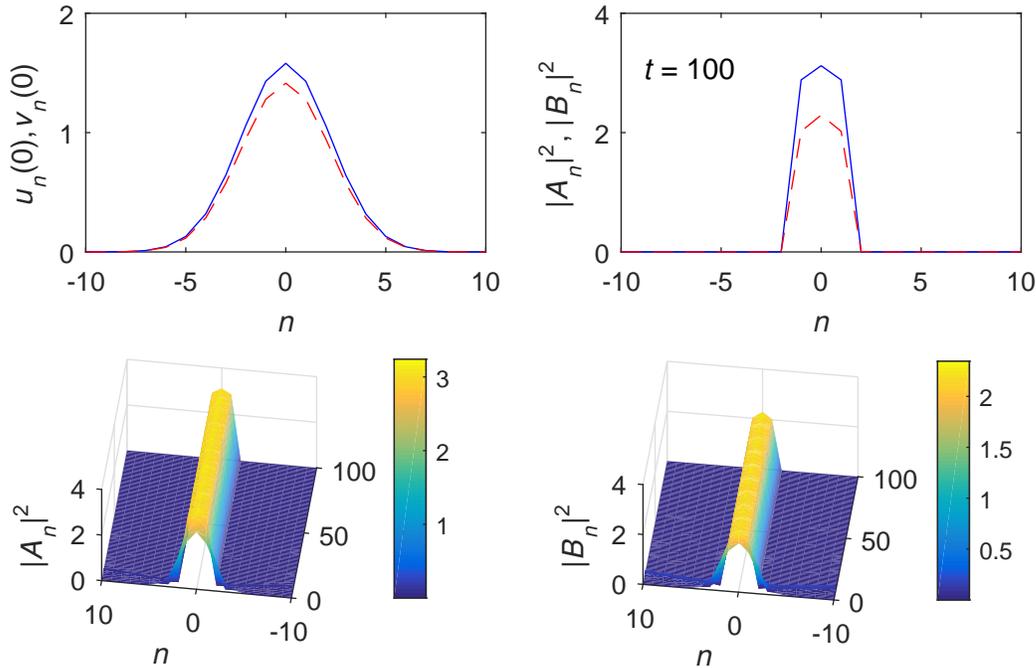}}\caption{(Color online) Top left panel. Initial Gaussian
profiles $u_{n}(0)=\sqrt{2.5}e^{-0.1n^{2}}$(solid line) and $v_{n}(0)=\sqrt{2.0}e^{-0.1n^{2}}$
(dash line) and at the right, the amplitudes at $t=100$ which are
practically three-sites compactons. Other parameters are fixed as
in Fig. \ref{BBg12-1}. Bottom plots show the transient amplitudes
along the time interval. To eliminate the matter radiation generated during the time evolution, the absorbing boundary conditions with the function $\xi_n$ as in~(\ref{bdiss}) with $\eta=10$, was used.}
\label{Addition4}
\end{figure}
By referring to the $^{41}$K-$^{87}$Rb BEC mixture, two Feshbach
resonances exist, one at  $35$G  and the other at $79$G~\cite{Thalhammer} . Considering the resonance at $B_{R}=35$G, we have $\Delta=(5.1\pm1.8)G$ with a background scattering length  $a_{bg}\approx 500 a_B $, where  $a_{B}=0.0529$nm is the Bohr radius. By assuming  $\Delta = 5.1$G, $\delta=0.08$G, $B_{1}=1.25$G, we have $\alpha=0.016,\, \beta=0.25$ and the first Fourier coefficients in Eq.~(\ref{Fourier}) are readily  estimated as  $a_0\approx -0.0154$, $a_1\approx 0.2538$, $a_2\approx 0.0317$, $a_3\approx 0.00396217$. From the numerical calculations of the averages in Eq. (\ref{averages}), one can check that all higher harmonics give negligibly contribution to the averaged  equations, so that, for any practical purpose, the effective inter-species modulation for the chosen parameters can be taken as in Eq.~(\ref{modulation}). Finally, we remark that  the optical lattice parameters can be fixed   according to the experiment~\cite{Catani}, e.g.,  $\lambda_{L}=1064$nm,  with  $V_{0}/E_{R}>10$, to guarantee the  validity of the tight-binding approximation~\cite{TS,ABKS,AKKS}.

In the SNM limit we expect compactons to be elementary excitations of the binary BEC system and as such they should emerge spontaneously from generic initial conditions. For  compactons detection in real  experiments, therefore,  one could start from generic localized matter wave-packet and observe their decomposition  into compacton modes localized on few optical lattice during the time evolution. An example of this is shown in  Fig.~\ref{Addition4} where we have numerically simulated a three-site binary  compacton formation starting from initial gaussian profiles for the two components. We see that the emitted radiation can be very small if the initial conditions are properly prepared (the numbers of atoms should be close to the ones theoretically predicted for compacton existence discussed in section $3$. For fixed parameters and qualitatively similar initial conditions, one should  observe delocalization transitions~\cite{Baizakov} occurring in one or in  both components,  that are  very sharp or quite broad, depending on how the initial condition were prepared, as the inter-species scattering length is varied across the threshold value (see section $4$).

\vskip .5cm

In conclusion, we have investigated in this paper the existence and stability of compactons and   quasi-compactons of binary BEC mixtures trapped in deep OL, induced by periodic time modulations of the inter-species scattering length. We showed that in the SNM limit exact stationary
compactons exist when the numbers of atoms in the two components are related  in a precise manner by a zero tunneling condition at the excitation edges. Slightly deviations from these conditions give rise to quasi-compactons, e.g. to solutions for which the zero tunneling condition and the localization on a compact is not exact but achieved dynamically through the coupling induced by the scattering length modulation. Stability properties of stationary compactons and
quasi-compactons have been extensively investigated both by linear analysis (for exact compactons)  and
by direct numerical integrations both of the averaged equations and of the original systems (compactons and quasi compactons cases). The comparison between analytical and numerical results are found in excellent agreement and confirm that compactons and
quasi-compactons  are very robust and  stable excitations. This particularly true for  inter-species scattering length that, for fixed parameters and initial conditions, are above the existence thresholds. Possible parameter designs and  experimental settings for BEC compactons observation was also discussed.

\section*{Acknowledgements}
M. S. acknowledges partial support from the Ministero dell'Istruzione,
dell'Universit\'a e della Ricerca (MIUR) through the PRIN (Programmi di
Ricerca Scientifica di Rilevante Interesse Nazionale) grant on
"Statistical Mechanics and Complexity": PRIN-2015-K7KK8L.
F.A, M.A.S.H., and B.A.U. acknowledges the support from the grant of MHE(Malaysia):
FRGS16-013-0512.\\

\end{document}